\documentclass[a4paper,12pt]{article}

\usepackage[english]{babel}
\usepackage[utf8x]{inputenc}
\usepackage[T1]{fontenc}

\usepackage[a4paper,top=2.5cm,bottom=2cm,left=2.5cm,right=2.5cm,marginparwidth=1.75cm]{geometry}

\usepackage{etoolbox}
\makeatletter
\patchcmd{\maketitle}{\@makefntext}{\fakecommand}{}{}
\patchcmd{\maketitle}{\rlap}{\hbox}{}{}
\patchcmd{\@maketitle}{\@author}{\hspace*{5pt}\@author}{}{}
\makeatother

\usepackage{amsmath}
\usepackage{graphicx}
\usepackage[colorlinks=true, allcolors=blue]{hyperref}
\usepackage{setspace}
\usepackage{apacite}
\usepackage{dsfont}
\usepackage{booktabs}
\usepackage[titletoc]{appendix}
\usepackage{tabularx} 
\usepackage{colortbl}
\usepackage{longtable}
\usepackage{threeparttablex} 
\usepackage{threeparttable}
\usepackage{chngcntr}
\usepackage{subcaption}
\usepackage[section]{placeins}
\usepackage{microtype}     
\usepackage{caption}        
\usepackage{rotating}
\usepackage{pdflscape}    
\usepackage{afterpage}
\usepackage{geometry}
\usepackage{color} 
\usepackage{float} 
\usepackage{lmodern}
\usepackage{hyperref}
\usepackage{subcaption}
\usepackage{bm}
\usepackage{amsfonts} 
\usepackage{makecell}
\usepackage{pifont}
\usepackage{cancel}
\usepackage[table]{xcolor}
\newcommand{\cmark}{\ding{51}}%

\usepackage[hang,flushmargin]{footmisc}

\DeclareMathOperator*{\E}{\mathbb{E}}

\newtheorem{cond}{Condition}

\newtheorem{definition}{Definition}

\newtheorem{prop}{Proposition}

\newtheorem{corollary}{Corollary}

\title{Treatment Effect Estimators as Weighted Outcomes} 

\author{Michael C. Knaus\thanks{University of T\"ubingen, Mohlstra{\ss}e 36, 72074 T\"ubingen, Germany. Michael C. Knaus is also affiliated with IZA, Bonn, \href{mailto:michael.knaus@uni-tuebingen.de}{michael.knaus@uni-tuebingen.de}. I thank participants of seminars at Aarhus, Bologna, CREST, D\"usseldorf, LISER, Mannheim, and St.~Gallen, of EuroCIM 2024 and VfS 2024, and in particular Oliver Dukes, Noah Greifer, Phillip Heiler, Michael Lechner, Dor Leventer, David Preinerstorfer, Christoph Rothe, Tymon S\l{}oczy\'nski, Anthony Strittmatter, Linbo Wang and Jose Zubizarreta for valuable comments improving the paper. I thank Henri Pfleiderer for assisting with the implementation of the R package. The usual disclaimer applies.}}

\date{\today}

\begin{document}
\maketitle

\onehalfspacing

\begin{abstract}
Estimators that weight observed outcomes to form effect estimates have a long tradition. Their outcome weights are widely used in established procedures, such as checking covariate balance, characterizing target populations, or detecting and managing extreme weights.
This paper introduces a general framework for deriving such outcome weights. It establishes when and how numerical equivalence between an original estimator representation as moment condition and a unique weighted representation can be obtained. The framework is applied to derive novel outcome weights for the six seminal instances of double machine learning and generalized random forests, while recovering existing results for other estimators as special cases.
The analysis highlights that implementation choices determine (i) the availability of outcome weights and (ii) their properties. 
Notably, standard implementations of partially linear regression-based estimators, like causal forests, employ outcome weights that do not sum to (minus) one in the (un)treated group, not fulfilling a property often considered desirable.
\bigskip


\end{abstract}

\newpage
\doublespacing

\section{Introduction}

Estimating the effect of treatment $D_i$ on outcome $Y_i$ is a common goal in causal inference. 
A variety of estimators is available for estimating different target parameters, after arguing for their identification within a particular research design \cite<see, e.g.~reviews by>{Imbens2009,Athey2017,Abadie2018EconometricEvaluation,Imbens2024CausalSciences}. Many of these estimators are a ``white box'' in the sense that they document how the sample is processed to obtain an effect estimate. Parametric regressions come with familiar coefficient outputs and other popular estimators have a representation as linear combination of observed outcomes:
\begin{align} \label{eq:general-w}
\hat{\tau} &= \sum_{i=1}^N \omega_i Y_i = \underbrace{\bm{\omega'}}_{1 \times N} \underbrace{\bm{Y}}_{N \times 1}
\end{align}
where $\omega_i$ represents the weight assigned to the outcome of observation $i$ in estimating $\hat{\tau}$.
Structure \eqref{eq:general-w} is most prominent in the literature on propensity score matching/weighting \cite<e.g.>{Imbens2015CausalSciences}, balancing estimators \cite<e.g.>{Ben-Michael2021TheInference}, and synthetic controls \cite<e.g.>{Abadie2021UsingAspects}. Furthermore, it is discussed for estimators of the local average treatment effect \cite<e.g.>{Imbens1997EstimatingModels,Abadie2003SemiparametricModels,Soczynski2024AbadiesEffect} as well as for linear regression \cite<e.g.>{Imbens2015MatchingExamples,Chattopadhyay2023OnInference}. 

Outcome weights $\omega_i$ have established use cases, such as: (i) \textit{covariate balancing checks} assessing internal validity in experimental and observational studies \cite<e.g.>{Rosenbaum1984ReducingScore,Rosenbaum1985ConstructingScore},
(ii) \textit{target population characterization} investigating external validity in IV settings \cite<e.g.>{Abadie2003SemiparametricModels}, or when using OLS \cite{Chattopadhyay2023OnInference}, (iii) \textit{extrapolation diagnostics} for estimators that could use negative weights \cite{Chattopadhyay2023OnInference}, (iv) \textit{finite sample estimator stabilization} by normalizing weights \cite<e.g.>{Hajek1971CommentOne}, or by trimming extreme weights \cite<e.g.>{Lechner2019PracticalEstimation}, (v) \textit{variance estimation} \cite<e.g.>[Ch. 19]{Imbens2015CausalSciences}.

In contrast, recent estimators integrating supervised machine learning into the estimation process \cite<see>[for a textbook]{Chernozhukov2024AppliedAI} can be considered as ``grey box''. Their multi-step algorithms are transparent and their theoretical properties are well-understood. However, neither coefficients nor outcome weights are currently available to interrogate how these steps jointly process a concrete sample within a concrete implementation. 
A key take-away of the analysis below is that at least outcome weights can be available for such multi-step estimators.

This paper introduces a simple but general framework to derive and analyze outcome weights of form \eqref{eq:general-w}.
We establish conditions for a numerical equivalence between an original estimator representation as moment condition and a unique weighted representation. 
The framework is applied to derive novel outcome weights for the six seminal instances of double machine learning \cite{Chernozhukov2018} and generalized random forest \cite{Athey2017a}.
Knowing the closed-form of the outcome weights has the immediate practical benefit that they can be plugged into established routines for classic weighting estimators. For example, covariate balancing can now be assessed for conditional average treatment effects estimated by causal forest. A second benefit is that the framework naturally allows to investigate basic properties of the weights. In particular, it highlights that implementation decisions control whether outcome weights of the treated sum up to one and of the untreated to minus one. Such weights are often considered as intuitive, reasonable and desirable in the literature \cite<e.g.>{Imbens2015CausalSciences,Soczynski2024AbadiesEffect}. However, the new framework reveals that estimators building on partially linear regression do not satisfy this property in standard implementations. 

The paper makes several contributions: (i) It introduces the first general framework to derive outcome weights; (ii) Its application to causal machine learning estimators yields novel outcome weights and provides a blueprint for applying the framework to other estimators; (iii) It illustrates how the new closed-form expressions enable established diagnostic tools from the weighting literature to be integrated off-the-shelf into causal machine learning applications; (iv) The theoretical results about conditions ensuring desirable estimator properties inform implementation decisions and complement the high-level conditions provided in asymptotic analyses; (v) The paper provides an additional piece in the continuing effort to blur the line between outcome weighting and outcome regression methods. \citeA{Bruns-Smith2023AugmentedRegression} show how weighting estimators can be expressed as regression estimators. This paper goes in the opposite direction by showing how estimators involving flexible outcome regression can be expressed as weighting estimators; (vi) The accompanying R package \href{https://github.com/MCKnaus/OutcomeWeights}{\texttt{OutcomeWeights}} computes the weights presented in the paper for general use \cite{Knaus2024OutcomeWeights}. The presented applications rely on this package and can be replicated in a supplementary \href{https://hub.docker.com/repository/docker/mcknaus/outcome_weights/general}{Docker image}.

\subsection{Related literature}

Outcome weights in form of \eqref{eq:general-w} are leveraged as common structure of difference, weighting, subclassification, and matching estimators \cite[Ch. 19.4.]{Smith2005DoesEstimators,Huber2013,Imbens2015CausalSciences}. Similarly, the outcome weights are derived and used for ordinary and weighted least squares based estimators \cite{Kline2011Oaxaca-BlinderEstimator,Imbens2015MatchingExamples,Jakiela2021SimpleEffects,Chattopadhyay2023OnInference,Hazlett2024UnderstandingSolutions}, two-stage least squares (TSLS) \cite{Chattopadhyay2021OnInference}, and augmented inverse probability weighting implemented with (post-selection) OLS outcome regression \cite{Knaus2021ASkills,Chattopadhyay2023OnInference}. This paper shows that a broader class of estimators can have this structure with a particular focus on those incorporating flexible outcome regression, while covering the results in the literature as special cases.

Other types of weights are prominent in the causal inference literature but distinct from the \textit{outcome weights} pursued in this paper. First, \textit{balancing weights} are the result of a tailored optimization problem to achieve covariate balancing of some prespecified form \cite{Graham2012InverseData,Hainmueller2012EntropyStudies,Imai2014CovariateScore,Zubizarreta2015StableData,Zhao2019CovariateFunctions,Kallus2020GeneralizedInference,Armstrong2021Finite-sampleUnconfoundedness,Heiler2022EfficientEffect}. Thus, balancing weights are an explicit part of such balancing estimators. While balancing weights are a special case of outcome weights, this paper focuses on estimators where the outcome weights play no explicit role but are implicit in the common characterization of the estimation procedure. \citeA{Chattopadhyay2023OnInference} call such weights ``implied weights'' in the context of OLS.
Second, \textit{effect weights} are central to understanding the estimand targeted by a given estimator. The pursued structures in this literature are variations of $\E[w(X_i)\tau(X_i)]$ where $w(X_i)$ is the weight an estimator assigns to the conditional treatment effect $\tau(X_i)$ in expectation. Effect weights are derived under different identifying and functional form assumptions for
OLS \cite{Angrist1998EstimatingApplicants,Angrist1999EmpiricalEconomics, Humphreys2009BoundsProbabilities, Aronow2016DoesEffects, Goldsmith-Pinkham2021ContaminationRegressions, Soczynski2022InterpretingWeights}, TSLS \cite{Imbens1994IdentificationEffects, Angrist1995Two-stageIntensity,Heckman2005StructuralEvaluation,Soczynski2020WhenLATEb,Blandhol2022WhenLate}, two-way fixed-effects \cite<see>[for overviews] {deChaisemartin2023Two-waySurvey,Roth2023WhatsLiterature}, regression discontinuity estimators \cite{Lee2010RegressionEconomics}, and panel estimators \cite{Chernozhukov2013AverageModels}.
The main difference between outcome weights and effect weights is that the former apply to observed outcomes and numerically reproduce the estimate without further assumptions, while the latter weight inherently unobservable effects and usually reproduce the estimate only in expectation. Both types of weights have their established use cases and are therefore complementary.

A small but growing body of literature provides nuance to the common notion that it is preferable for the outcome weights to sum to (minus) one within treatment groups. \citeA{Doudchenko2016BalancingSynthesis} and \citeA{Breitung2024AlternativeDesigns} challenge this view for synthetic control estimators, and  \citeA{Khan2023AdaptiveEstimation} for average treatment effect estimation. \citeA{Soczynski2024AbadiesEffect} note that some Abadie's \citeyear{Abadie2003SemiparametricModels} $\kappa$ estimators are intermediate cases between normalized and unnormalized estimators, e.g. with treated weights summing to one but untreated weights not to summing minus one. This paper adds the observation that partially linear regression based estimators usually produce weights that overall sum up to zero but not to (minus) one for (un)treated.

Finally, the paper adds to recent works establishing numeric equivalences between different estimator representations \cite{Bruns-Smith2023AugmentedRegression} or estimators \cite{Soczynski2023CovariateEffects,Soczynski2024AbadiesEffect} for conceptual and/or practical insights.

\section{A general framework to derive outcome weights} \label{sec:general-framework}

\subsection{Notation}

The estimators under consideration require access to data with $N$ observations indexed by $i = 1,...,N$. The data includes a binary treatment $D_i$, an outcome $Y_i$, covariates $\bm{X_i}$, and an optional binary instrument $Z_i$, all collected in $O_i = (D_i, \bm{X_i}', Y_i, Z_i)'$. The empirical mean of a variable $A_i$ is represented as $\E_N[A_i] = N^{-1}\sum_{i=1}^N A_i$. 

Many results are stated in matrix notation where bold letters describe vectors or matrices of variables.
$\bm{I_k}$ denotes the identity matrix of dimension $k$, $\bm{0_k}$ and $\bm{1_k}$ represent column vectors of length $k$ containing zeros and ones, respectively.

\subsection{Pseudo-IV estimators}

This paper focuses on estimators falling into the class of pseudo-IV estimators (PIVE):
\begin{definition}\label{def:pive} (pseudo-IV estimators) \\
Define the class of pseudo-IV estimators (PIVE) as estimators solving an empirical moment condition of the form
\begin{equation} \label{eq:pive}
\mathbb{E}_N \left[(\tilde{Y}_i - \hat{\tau} \tilde{D}_i) \tilde{Z}_i  \right] = 0  
\end{equation}
with 
\begin{itemize} \singlespacing
\item $\tilde{Y}_i = f_Y(O_i;\hat{\eta}_i^Y)$: scalar pseudo-outcome 
\item $\tilde{D}_i = f_D(O_i;\hat{\eta}_i^D)$: scalar pseudo-treatment
\item $\tilde{Z}_i = f_Z(O_i;\hat{\eta}_i^Z)$: scalar pseudo-instrument
\end{itemize}
where $\hat{\eta}_i^Y$, $\hat{\eta}_i^D$ and $\hat{\eta}_i^Z$ are optional nuisance parameters.
\end{definition}

Note that the PIVE representation is neither a unique, nor the most compact representation of an estimator. For example, a representation using a linear score $\mathbb{E}_N \left[\psi_i^a - \hat{\tau} \psi_i^b \right] = 0$ with $\psi_i^a =\tilde{Y}_i \tilde{Z}_i$ and $\psi_i^b =\tilde{D}_i \tilde{Z}_i$ would be equivalent, or vice versa any estimator with a linear score can be written as PIVE with $\tilde{Y}_i = \psi_i^a$, $\tilde{D}_i = \psi_i^b$, and $\tilde{Z}_i = 1$. However, the PIVE structure is essential for the goal of this paper. In particular, separating the pseudo-outcome from the pseudo-instrument makes the derivation and analysis of the outcome weights tractable.

\textit{Example (OLS):} We use the canonical OLS estimator as a running example to illustrate the general results throughout the paper. Consider a linear outcome model $Y_i = \tau D_i + \bm{X_i'\beta} + \varepsilon$. The OLS estimator for $\tau$ can be expressed as PIVE using the residual-on-residual regression representation of the Frisch-Waugh-Lovell Theorem:

\begin{equation} \label{eq:mom-ols}
\mathbb{E}_N \big[ \{\underbrace{Y_i - \bm{X_i^{'}} \hat{\beta}_{\bm{Y}  \sim \bm{X}}}_{=: \tilde{Y}_i^{ols}} - \hat{\tau}^{ols} \underbrace{[D_i - \bm{X_i^{'}} \hat{\beta}_{\bm{D}  \sim \bm{X}}]}_{=: \tilde{D}_i^{ols}} \} \underbrace{[D_i - \bm{X_i^{'}} \hat{\beta}_{\bm{D}  \sim \bm{X}}]}_{=: \tilde{Z}_i^{ols}} \big] = 0   
\end{equation}
where $\hat{\beta}_{\bm{Y} \sim \bm{X}} := \bm{(X'X)^{-1} X'Y}$ and $\hat{\beta}_{\bm{D} \sim \bm{X}} := \bm{(X'X)^{-1} X'D}$ such that the pseudo-outcome is the outcome residual, and both pseudo-treatment and -instrument are the treatment residual.

\subsection{Outcome weights of pseudo-IV estimators}

Solving Equation \ref{eq:pive} leads to parameter estimate
\begin{equation}
\label{eq:pive1}
\hat{\tau} = \frac{\E_N \left[\tilde{Z}_i \tilde{Y}_i\right]}{\E_N \left[\tilde{Z}_i \tilde{D}_i\right]} = (\bm{\tilde{Z}'\tilde{D}})^{-1}\bm{\tilde{Z}'} \bm{\tilde{Y}}.
\end{equation}
Now assume that the pseudo-outcome vector can be obtained by multiplying a unique $N \times N$ transformation matrix $\bm{T}$ with the outcome vector, i.e.~$\bm{TY} = \bm{\tilde{Y}}$. Then, Equation \ref{eq:pive1} can be written in the form of Equation \ref{eq:general-w}
\begin{equation}
\label{eq:pive2}
\hat{\tau} = \underbrace{(\bm{\tilde{Z}'\tilde{D}})^{-1}\bm{\tilde{Z}'T}}_{\bm{\omega'}} \bm{Y} = \bm{\omega' Y}
\end{equation}
leading to a core result of the paper:
\begin{prop}\label{prop:iw-pive} (outcome weights of PIVE) \\
The outcome weights of a PIVE in the sense of Definition \ref{def:pive} have closed-form
\begin{equation}
\label{eq:general-pl}
\bm{\omega'} = (\bm{\tilde{Z}'\tilde{D}})^{-1}\bm{\tilde{Z}'T}
\end{equation}
if $\bm{\tilde{Z}'\tilde{D}} \neq 0$ and a unique transformation matrix $\bm{T}$ exists such that $\bm{TY} = \bm{\tilde{Y}}$.
\end{prop}

This simple result is constructive because it motivates a two step procedure to derive outcome weights:
\begin{enumerate}
\item Express the estimator as PIVE.
\item Find transformation matrix $\bm{T}$ $\Rightarrow$ $\bm{\omega}$ has closed-form.
\end{enumerate}
These steps are illustrated below for a variety of estimators. However, the procedure is general and could be pursued for any other estimator fitting into the PIVE structure.

\textit{Example (OLS) continued:} The solution of the residual-on-residual regression \eqref{eq:mom-ols} in form of Equation \ref{eq:general-pl} is 
\begin{align} \label{eq:ols-exmpl}
\hat{\tau}^{ols} = \overbrace{(\underbrace{\bm{\hat{E}'}}_{\bm{\tilde{Z}^{ols'}}}\underbrace{\bm{\hat{E}}}_{\bm{\tilde{D}^{ols}}})^{-1} \underbrace{\bm{\hat{E'}}}_{\bm{\tilde{Z}^{ols'}}} \underbrace{\bm{M_X}}_{\bm{T}^{ols}}}^{\bm{\omega}^{ols'}} \bm{Y}
\end{align}
where we use the projection matrix $\bm{P_X} := \bm{X(X'X)^{-1} X'}$ to define the residual maker matrix $\bm{M_X} := \bm{I_N - P_X}$, and the treatment residual vector $\bm{\hat{E}} := \bm{M_X D}$.
The residual maker matrix is therefore the outcome transformation matrix of OLS and $\bm{\omega}^{ols'} = (\bm{\hat{E}' \hat{E}})^{-1}\bm{\hat{E}' M_X}$ is the outcome weights vector.\footnote{We deliberately do not use that $\bm{M_X}$ is idempotent for illustration purposes but note that also the identity matrix would by a suitable transformation matrix in \eqref{eq:ols-exmpl}.}

\section{Outcome weights of concrete pseudo-IV estimators} \label{sec:ow-concrete}

This section leverages the new framework to provide the first characterization of outcome weights for six seminal instances within the double machine learning (DML) and generalized random forest (GRF) frameworks (marked with $^*$), while also recovering existing results for eight other estimators:
\begin{itemize} \singlespacing
\item \textbf{IF$^*$: Instrumental forest} \cite{Athey2017a} 
\item PLR-IV$^*$: Partially linear regression with IV \cite{Chernozhukov2018}
\item TSLS: Two stage least squares
\item Wald: Wald estimator \cite{Wald1940TheError}
\item CF$^*$: Causal forest \cite{Athey2017a}
\item PLR$^*$: Partially linear regression \cite{Robinson1988Root-n-consistentRegression,Chernozhukov2018}
\item OLS: Ordinary least squares
\item DiM: Difference in means
\item \textbf{AIPW$^*$: Augmented inverse probability weighting} \cite{Robins1995SemiparametricData,Chernozhukov2018}
\item RA: Regression adjustment \cite<e.g.~discussed by>{Imbens2004NonparametricReview}
\item IPW: Inverse probability weighting \cite{Horvitz1952}
\item \textbf{Wald-AIPW$^*$: Wald type AIPW} \cite{Tan2006RegressionVariables,Chernozhukov2018}
\item Wald-RA: Wald type regression adjustment \cite{Tan2006RegressionVariables}
\item Wald-IPW: Wald type inverse probability weighting \cite{Tan2006RegressionVariables}
\end{itemize}

Conveniently it suffices to analyze the three estimators in bold letters - IF, AIPW and Wald-AIPW - because their subsequent estimators follow as special cases (see Figure \ref{fig:estimators} for a graphical illustration). Each estimator is typically used at an intersection of three research designs (randomized controlled trials, unconfoundedness or instrumental variables), two aggregation levels (average or conditional effects), and three outcome model assumptions (none, partially linear, or linear models). Appendix \ref{sec:app-estimators} summarizes the causal parameters and settings for which each estimator is usually applied. However, the main text ignores definition, identification and interpretation issues concentrating on the mechanics of the estimators.

\subsection{Nuisance parameters}

\subsubsection{Definitions}

The considered estimators require a variety of nuisance parameters in the form of approximated conditional expectations:
\begin{align}
\hat{Y}_i &:= \hat{\E}[Y_i|\bm{X_i}] & \hat{D}_i &:= \hat{\E}[D_i|\bm{X_i}] \nonumber \\
\hat{Y}^d_{0,i} &:= \hat{\E}[Y_i|D_i = 0,\bm{X_i}] & \hat{D}_{0,i}^z &:= \hat{\E}[D_i|Z_i = 0,\bm{X_i}] \nonumber \\    
\hat{Y}^d_{1,i} &:= \hat{\E}[Y_i|D_i = 1,\bm{X_i}] & \hat{D}_{1,i}^z &:= \hat{\E}[D_i|Z_i = 1,\bm{X_i}] \\    
\hat{Y}^z_{0,i} &:= \hat{\E}[Y_i|Z_i = 0,\bm{X_i}] & \hat{Z}_i &:= \hat{\E}[Z_i|\bm{X_i}] \nonumber \\    
\hat{Y}^z_{1,i} &:= \hat{\E}[Y_i|Z_i = 1,\bm{X_i}] &  \nonumber
\end{align}
Furthermore, define the inverse probability weights of the treated $\lambda_{1,i}^{ipw} := 
D_i / \hat{D}_i$ and of the untreated $\lambda_{0,i}^{ipw} := 
(1-D_i) / (1-\hat{D}_i)$.\footnote{We use $\lambda$ to remind us that these weights are on a different scale than the $\omega$ weights. Using the corresponding $\omega_{d,i}^{ipw} := \lambda_{d,i}^{ipw} / N$ definition would unnecessarily complicate notation below.} Similarly, define the instrument inverse probability weights as $\lambda_{1,i}^{ipw,z} := Z_i / \hat{Z}_i$ and $\lambda_{0,i}^{ipw,z} := (1-Z_i) / (1-\hat{Z}_i)$.

\subsubsection{A crucial building block: Smoothers}

The literature knows numerous regression methods to estimate the outcome nuisance parameters $\hat{Y}_i$, $\hat{Y}_{d,i}^d$, and $\hat{Y}_{z,i}^z$. However, the class of smoothers \cite<see e.g.>[Ch. 2-3]{Hastie1990GeneralizedModels} turns out to be crucial for the purpose of this paper. Smoothers produce outcome predictions by weighting/smoothing observed outcomes
\begin{equation}
\hat{Y}_i = \sum_{j=1}^N s_{i\leftarrow j} Y_j    
\end{equation}
where the \textit{smoother weight} $s_{i\leftarrow j}$ represents the contribution of unit $j$'s outcome to the prediction of unit $i$.\footnote{The arrow notation is adapted from \citeA{Lin2022OnEffect}.} Define also the \textit{smoother vector} for the outcome prediction of unit $i$ by $\bm{s}_i = (s_{i\leftarrow 1},...,s_{i\leftarrow N})'$ and the $N \times N$ \textit{smoother matrix} $\bm{S} = [\bm{s}_1~...~\bm{s}_N]'$ such that $\bm{s}_i' \bm{Y} = \hat{Y}_i$ and $\bm{S Y = \hat{Y}}$.

The smoother weights in this paper are explicitly allowed to depend on the outcomes (adaptive smoother) and on random components (random smoother), i.e.~$\bm{s}_i := \bm{s}_i(\bm{X_i};\bm{X},\bm{Y},\epsilon_s)$.\footnote{The categorization of smoothers is inspired by \citeA{Curth2024WhySmoothers}.} This covers for example (post-selection) OLS,
ridge, spline and kernel (ridge) regressions, regression trees, random forests or boosted trees with data-driven hyperparameter tuning (see Appendix \ref{sec:app-smoothers} for further discussion). However, the numerical equivalences established below require the mere existence of a smoother matrix:
\begin{cond}\label{cond:smoothers} (smoother matrix) \\
There exists a unique smoother matrix creating the outcome nuisance vectors if multiplied with the outcome vector:

\noindent (C1a) $\bm{SY} = \bm{\hat{Y}}$

\noindent (C1b) $\bm{S_0^d Y} = \bm{\hat{Y}_0^d}$ and $\bm{S_1^d Y} = \bm{\hat{Y}_1^d}$ 

\noindent (C1c) $\bm{S_0^z Y} = \bm{\hat{Y}_0^z}$ and $\bm{S_1^z Y} = \bm{\hat{Y}_1^z}$ 
\end{cond}

\textit{Example (OLS) continued:} The projection matrix is arguably the most prominent smoother matrix producing fitted values of an OLS regression as $\bm{\underbrace{P_X}_{S^{ols}} Y} = \bm{\hat{Y}^{ols}}$.

\subsection{Concrete outcome weights} \label{sec:iw}

\subsubsection{Instrumental forest and its special cases} \label{sec:iw-if}

The instrumental forest (IF) of \citeA{Athey2017a} runs an $\bm{x}$-specific weighted partially linear IV regression
\begin{equation} \label{eq:mom-cf}
\mathbb{E}_N \big[ \{\underbrace{Y_i - \hat{Y}_i}_{=: \tilde{Y}_i^{if}} - \hat{\tau}^{if}(\bm{x}) \underbrace{[D_i - \hat{D}_i]}_{=: \tilde{D}_i^{if}} \} \underbrace{[Z_i - \hat{Z}_i] \alpha_i^{if}(\bm{x})}_{=:\tilde{Z}_i^{if}} \big] = 0   
\end{equation}
where the $\bm{x}$-specific weights $\alpha^{if}(\bm{x})$ are obtained by the tailored splitting criterion described in \citeA{Athey2017a} and can be extracted via the \texttt{get\_forest\_weights()} function of their \texttt{grf} R package \cite{Tibshirani2024Grf:Forests}. The solution in the form of Equation \ref{eq:pive1} is
\begin{equation} \label{eq:if}
\hat{\tau}^{if}(\bm{x}) = (\bm{\hat{R}'}diag(\bm{\alpha^{if}(\bm{x})})\bm{\hat{V}})^{-1} \bm{\hat{R}'} diag(\bm{\alpha^{if}(\bm{x})}) \bm{\hat{U}}
\end{equation}
where $\bm{\hat{R}} = \bm{Z} - \bm{\hat{Z}}$, $\bm{\hat{V}} = \bm{D} - \bm{\hat{D}}$ and  $\bm{\hat{U}} = \bm{Y} - \bm{\hat{Y}}$ are the instrument, treatment and outcome residual vectors, respectively. The PIVE structure is therefore established. The next step is to understand whether the pseudo-outcome can be obtained using a transformation matrix. This is only possible if a smoother is applied to obtain the outcome predictions, i.e.~Condition \ref{cond:smoothers}a holds such that $\bm{\hat{U}} =  \bm{Y} - \bm{S Y} = (\bm{I_N} - \bm{S}) \bm{Y}$ and 
\begin{equation} \label{eq:if2}
\hat{\tau}^{if}(\bm{x}) = \underbrace{(\bm{\hat{R}'}diag(\bm{\alpha^{if}(\bm{x})})\bm{\hat{V}})^{-1} \bm{\hat{R}'}diag(\bm{\alpha^{if}(\bm{x})}) \overbrace{(\bm{I_N} - \bm{S})}^{\bm{T^{if}}}}_{\bm{\omega^{if'}}} \bm{Y}.
\end{equation}
The transformation matrix of IF can therefore be considered as a generalized residual maker matrix. Equation \ref{eq:if2} contains then the first concrete case of Proposition \ref{prop:iw-pive}:
\begin{corollary}\label{cor:if} (outcome weights of instrumental forest) \\ 
Under Condition \ref{cond:smoothers}a such that the outcome predictions can be written as $\bm{S Y} = \bm{\hat{Y}}$,
the outcome weights of instrumental forests take the form
\begin{equation} \label{eq:iw-if}
\bm{\omega}^{if}(\bm{x})\bm{'} = (\bm{\hat{R}'}diag(\bm{\alpha^{if}(\bm{x})})\bm{\hat{V}})^{-1} \bm{\hat{R}'}diag(\bm{\alpha^{if}(\bm{x})}) (\bm{I_N} - \bm{S}).
\end{equation}
\end{corollary}

Table \ref{tab:if-paths} compactly shows how seven other estimators (in light gray) follow as special cases of IF. Starting from the dark gray row, we can follow an upward path to the Wald estimator or a downward path to DiM. The white rows between the gray rows document the modifications needed to recover the next estimator. For example, moving from IF to CF uses treatment residuals instead of instrument residuals and the CF specific weights $\bm{\alpha^{cf}}$ in the pseudo-instrument, while pseudo-treatment and transformation matrix remain unchanged. Similarly setting the weights to one recovers PLR from CF and PLR-IV from IF. Continuing the paths up- and downwards replaces the generic predictions with linear projections to recover TSLS and OLS, respectively. Finally, using the projection matrix of a constant recovers Wald estimator and DiM.

\begin{table}[h!]
\captionsetup{skip=-5pt}
    \caption{IF pseudo-variables and transformation matrices}
    \label{tab:if-paths}
    \centering
    \onehalfspacing
    \begin{threeparttable}
\[
\begin{array}{|l|c|c|c|}
    \hline
      & \bm{\tilde{Z}'} & \bm{\tilde{D}} & \bm{T} \\
    \hline
\rowcolor{gray!30} \text{Wald} & \bm{Z'}\bm{M_{1_N}} & \bm{M_{1_N}} \bm{D} & \bm{M_{1_N}} \\
        & \uparrow \bm{P_X} = \bm{P_{1_N}} \uparrow& \uparrow \bm{P_X} = \bm{P_{1_N}} \uparrow & \uparrow \bm{P_X} = \bm{P_{1_N}} \uparrow \\
\rowcolor{gray!30} \text{TSLS} &  \bm{Z'}\bm{M_X} & \bm{M_X}\bm{D} & \bm{M_X} \\
     & \uparrow \bm{\hat{Z}} = \bm{P_XZ} \uparrow& \uparrow \bm{\hat{D}} = \bm{P_XD} \uparrow & \uparrow \bm{\hat{Y}} = \bm{P_XY} \uparrow \\
\rowcolor{gray!30} \text{PLR-IV} & \bm{\hat{R}'}  &\bm{\hat{V}} & (\bm{I_N} - \bm{S}) \\
     & \uparrow \bm{\alpha^{if}} = \bm{1_N} \uparrow& = & = \\
\rowcolor{gray!65} \text{IF} & \bm{\hat{R}'}diag(\bm{\alpha^{if}}(\bm{x})) &\bm{\hat{V}} & (\bm{I_N} - \bm{S}) \\
     & \downarrow \bm{\hat{R}} = \bm{\hat{V}}~\&~\bm{\alpha^{if}} = \bm{\alpha^{cf}} \downarrow& = & = \\   
\rowcolor{gray!30} \text{CF} & \bm{\hat{V}'}diag(\bm{\alpha^{cf}}(\bm{x})) &\bm{\hat{V}} & (\bm{I_N} - \bm{S}) \\
     & \downarrow  \bm{\alpha^{cf}} = \bm{1_N} \downarrow& = & = \\         
\rowcolor{gray!30} \text{PLR} & \bm{\hat{V}'} &\bm{\hat{V}} & (\bm{I_N} - \bm{S}) \\
    & \downarrow \bm{\hat{D}} = \bm{P_XD} \downarrow& \downarrow \bm{\hat{D}} = \bm{P_XD} \downarrow & \downarrow \bm{\hat{Y}} = \bm{P_XY} \downarrow \\    
\rowcolor{gray!30} \text{OLS} &  \bm{D'}\bm{M_X} & \bm{M_X}\bm{D} & \bm{M_X} \\
& \downarrow \bm{P_X} = \bm{P_{1_N}} \downarrow& \downarrow \bm{P_X} = \bm{P_{1_N}} \downarrow & \downarrow \bm{P_X} = \bm{P_{1_N}} \downarrow \\
\rowcolor{gray!30} \text{DiM} & \bm{D'}\bm{M_{1_N}} & \bm{M_{1_N}} \bm{D} & \bm{M_{1_N}} \\
\hline
\end{array}
\]
\begin{tablenotes} \small \item \textit{Note:} Starting from the darkest row and following the arrows, the table shows how estimators follow as special cases by imposing restrictions in the white rows. \end{tablenotes}  
\end{threeparttable}
\end{table}

\textit{Computational remark:} The original implementation in the \texttt{grf} package applies a constant in the weighted residual-on-residual regression. This complicates notation but Appendix \ref{sec:app-grf-cf} provides the details how numerical equivalence between original output of \texttt{grf} and the weighted representation is obtained in the \texttt{OutcomeWeights} package.

\subsubsection{Augmented inverse probability weighting and its special cases} \label{sec:aipw-ow}

Augmented inverse probability weighting (AIPW) is developed in a series of papers \cite<e.g.>{Robins1994,Robins1995AnalysisData,Rotnitzky1998SemiparametricNonresponse,Chernozhukov2018}. AIPW is a PIVE with empirical moment condition
\vspace{-0.2cm}
\begin{equation} \label{eq:mom-aipw}
\mathbb{E}_N \Bigg[\Big\{ \underbrace{\hat{Y}^d_{1,i} - \hat{Y}^d_{0,i} + \lambda_{1,i}^{ipw} (Y_i - \hat{Y}^d_{1,i})  - \lambda_{0,i}^{ipw}(Y_i - \hat{Y}^d_{0,i})}_{=:\tilde{Y}_i^{aipw}} - \hat{\tau}^{aipw} \underbrace{1}_{=:\tilde{D}_i^{aipw}} \Big\} \underbrace{1}_{=:\tilde{Z}_i^{aipw}} \Bigg] = 0
\end{equation}
and vector form
\begin{equation} \label{eq:aipw-4}
\hat{\tau}^{aipw} = (\bm{1_N'}\bm{1_N})^{-1} \bm{1_N'}[ \bm{\hat{Y}^d_1} - \bm{\hat{Y}^d_0} +  diag(\bm{\lambda_1^{ipw}}) (\bm{Y} - \bm{\hat{Y}^d_1}) -  diag(\bm{\lambda_0^{ipw}}) (\bm{Y} - \bm{\hat{Y}^d_0}) ]. 
\end{equation}

The next step is to provide the transformation matrix. This is possible under Condition \ref{cond:smoothers}b that the outcome predictions are obtained by smoothers such that $\bm{S^d_d Y} = \bm{\hat{Y}^d_d}$. Plugging this into \eqref{eq:aipw-4} and rearranging delivers the transformation matrix
\begin{equation}
\hat{\tau}^{aipw} = N^{-1} \bm{1_N'} [ \underbrace{ \bm{S^d_1} - \bm{S^d_0} + diag(\bm{\lambda_{1}^{ipw}}) (\bm{I_N} - \bm{S^d_1}) -  diag(\bm{\lambda_{0}^{ipw}}) (\bm{I_N} - \bm{S^d_0}) }_{=:\bm{T^{aipw}}} ] \bm{Y}    
\end{equation}
and leads to the following result:\footnote{The AIPW implementation of the \texttt{grf} package uses an alternative moment condition. It is equivalent to \eqref{eq:mom-aipw} in expectation but uses different nuisance parameters and therefore differs numerically. However, also the outcome weights of this variant can be obtained as shown in \ref{sec:app-grf-aipw}.}
\begin{corollary}\label{cor:aipw} (outcome weights of AIPW) \\ 
Under Condition \ref{cond:smoothers}b such that the treatment specific outcome predictions can be written as  $\bm{S^d_1 Y} = \bm{\hat{Y}^d_1}$ and $\bm{S^d_0 Y} = \bm{\hat{Y}^d_0}$, the outcome weights of AIPW take the form
\begin{align} \label{eq:iw-aipw}
\bm{\omega^{aipw'}} & = N^{-1} \bm{1_N'} [ \bm{S^d_1} - \bm{S^d_0} +  diag(\bm{\lambda_1^{ipw}}) (\bm{I_N} - \bm{S^d_1}) -  diag(\bm{\lambda_0^{ipw}}) (\bm{I_N} - \bm{S^d_0})]. 
\end{align}
\end{corollary}

Table \ref{tab:aipw-paths} shows how RA can be obtained by setting all IPW weights to zero. IPW is recovered by setting all entries of the smoother matrices to zero. 

\begin{table}[h!]
\captionsetup{skip=-5pt}
    \caption{AIPW pseudo-variables and transformation matrices}
    \label{tab:aipw-paths}
    \centering
    \onehalfspacing
    \begin{threeparttable}
\[
\begin{array}{|l|c|c|c|}
    \hline
      & \bm{\tilde{Z}'} & \bm{\tilde{D}} & \bm{T} \\
        \hline
\rowcolor{gray!30} \text{RA} & \bm{1_N'}  & \bm{1_N} & \bm{S^d_1} - \bm{S^d_0} \\
     &  = & = & \uparrow \bm{\lambda_{1}^{ipw}} = \bm{\lambda_{0}^{ipw}} = \bm{0_N} \uparrow \\
\rowcolor{gray!65} \text{AIPW} & \bm{1_N'} & \bm{1_N} & \bm{S^d_1} - \bm{S^d_0} + diag(\bm{\lambda_{1}^{ipw}}) (\bm{I_N} - \bm{S^d_1}) -  diag(\bm{\lambda_{0}^{ipw}}) (\bm{I_N} - \bm{S^d_0}) \\
     & = & = &  \downarrow \bm{S^d_1} = \bm{S^d_0} = \bm{0_{N \times N}} \downarrow \\   
\rowcolor{gray!30} \text{IPW} & \bm{1_N'} & \bm{1_N} & diag(\bm{\lambda_{1}^{ipw}} - \bm{\lambda_{0}^{ipw}}) \\
\hline
\end{array}
\]
\end{threeparttable}
\end{table}

\subsubsection{Wald-AIPW and its special cases} \label{sec:wald-aipw-ow}

\citeA{Tan2006RegressionVariables} propose an AIPW extension for the case of a binary instrument. This estimator has the same structure as the canonical \citeA{Wald1940TheError} estimator but applies AIPW to estimate reduced form and first stage, respectively. Following \citeA{Chernozhukov2018}, the Wald-AIPW empirical moment condition in the form of Equation \ref{eq:pive} reads
\begin{align}\label{eq:iv-ident}
\mathbb{E}_N &\Bigg[ \Bigg\{ \overbrace{ \hat{Y}^z_{1,i} - \hat{Y}^z_{0,i} + \lambda_{1,i}^{ipw,z} (Y_i - \hat{Y}^z_{1,i}) - \lambda_{0,i}^{ipw,z} (Y - \hat{Y}^z_{0,i}) }^{ \tilde{Y}_i^{iv-aipw}:=} \\
&\quad  - \hat{\tau}^{iv-aipw} \Bigg( \underbrace{\hat{D}_{1,i}^z - \hat{D}_{0,i}^z + \lambda_{1,i}^{ipw,z} (D_i - \hat{D}_{1,i}^z) - \lambda_{1,i}^{ipw,z} (D_i - \hat{D}_{0,i}^z)}_{=: \tilde{D}_i^{iv-aipw}} \Bigg) \Bigg\}  \underbrace{1}_{=:\tilde{Z}_i^{iv-aipw}} \Bigg]  = 0. \nonumber
\end{align}
and in the form of Equation \ref{eq:pive1} becomes
\begin{align}
\hat{\tau}^{iv-aipw} & = (\bm{1_N'}[ \bm{\hat{D}^z_1} - \bm{\hat{D}^z_0} +  diag(\bm{\lambda_1^{ipw,z}}) (\bm{D} - \bm{\hat{D}^z_1}) -  diag(\bm{\lambda_0^{ipw,z}}) (\bm{D} - \bm{\hat{D}^z_0}) ])^{-1} \\
& \quad \times \bm{1_N'}[ \bm{\hat{Y}^z_1} - \bm{\hat{Y}^z_0} +  diag(\bm{\lambda_1^{ipw,z}}) (\bm{Y} - \bm{\hat{Y}^z_1}) -  diag(\bm{\lambda_0^{ipw,z}}) (\bm{Y} - \bm{\hat{Y}^z_0}) ].  \nonumber  
\end{align}
Following similar steps as in Section \ref{sec:aipw-ow} establishes another special case of Proposition \ref{prop:iw-pive}:
\begin{corollary}\label{cor:w-aipw} (outcome weights of Wald-AIPW) \\ 
Under Condition \ref{cond:smoothers}c such that the instrument specific outcome predictions can be written as  $\bm{S^z_1 Y} = \bm{\hat{Y}^z_1}$ and $\bm{S^z_0 Y} = \bm{\hat{Y}^z_0}$, the outcome weights of Wald-AIPW take the form
\begin{align} \label{eq:iw-iv-aipw}
\bm{\omega^{iv-aipw'}} & = (\bm{1_N'}\overbrace{[ \bm{\hat{D}^z_1} - \bm{\hat{D}^z_0} +  diag(\bm{\lambda_1^{ipw,z}}) (\bm{D} - \bm{\hat{D}^z_1}) -  diag(\bm{\lambda_0^{ipw,z}}) (\bm{D} - \bm{\hat{D}^z_0}) ]}^{\bm{\tilde{D}^{iv-aipw }}:=})^{-1} \nonumber \\
& \quad \times \bm{1_N'} \underbrace{[ \bm{S^z_1} - \bm{S^z_0} +  diag(\bm{\lambda_1^{ipw,z}}) (\bm{I_N} - \bm{S^z_1}) -  diag(\bm{\lambda_0^{ipw,z}}) (\bm{I_N} - \bm{S^z_0})]}_{=: \bm{T^{iv-aipw}}}. 
\end{align}
\end{corollary}

Table \ref{tab:wald-aipw-paths} summarizes the involved manipulations to arrive at Wald-RA and -IPW applying similar transformations as for AIPW but for both reduced form and first stage.

\begin{table}[h!]
\captionsetup{skip=-5pt}
    \caption{Wald-AIPW pseudo-variables and transformation matrices}
    \label{tab:wald-aipw-paths}
    \centering
    \onehalfspacing
\begin{threeparttable}
\[
\begin{array}{|l|c|c|c|}
    \hline
      & \bm{\tilde{Z}'} & \bm{\tilde{D}} & \bm{T} \\
        \hline
\rowcolor{gray!30} \text{Wald-RA} & \bm{1_N'}  & \bm{\hat{D}^z_1} - \bm{\hat{D}^z_0} & \bm{S^z_1} - \bm{S^z_0} \\
     &  = & \uparrow \bm{\lambda_{1}^{ipw,z}} = \bm{\lambda_{0}^{ipw,z}} = \bm{0_N} \uparrow & \uparrow \bm{\lambda_{1}^{ipw,z}} = \bm{\lambda_{0}^{ipw,z}} = \bm{0_N} \uparrow \\
\rowcolor{gray!65} \text{Wald-AIPW} & \bm{1_N'} & \bm{\tilde{D}^{iv-aipw }} \text{ in \eqref{eq:iw-iv-aipw}} & \bm{\tilde{T}^{iv-aipw }} \text{ in \eqref{eq:iw-iv-aipw}} \\
     & = & \downarrow \bm{\hat{D}^z_1} = \bm{\hat{D}^z_0} \downarrow = \bm{0_N} &  \downarrow \bm{S^z_1} = \bm{S^z_0} = \bm{0_{N \times N}} \downarrow \\   
\rowcolor{gray!30} \text{Wald-IPW} & \bm{1_N'} & diag(\bm{\lambda_{1}^{ipw,z}} - \bm{\lambda_{0}^{ipw,z}}) & diag(\bm{\lambda_{1}^{ipw,z}} - \bm{\lambda_{0}^{ipw,z}}) \\
\hline
\end{array}
\]
\end{threeparttable}
\end{table}
\vspace{-.5cm}

\subsection{Consolidation}

This section provides the first characterization of outcome weights for IF, CF, PLR(-IV), and (Wald-)AIPW (the supplementary \href{https://mcknaus.github.io/assets/notebooks/outcome_weights/Theory%20in%20action.nb.html}{theory in action notebook} illustrates that the numerical equivalences hold also in practice).
The results highlight that the availability of outcome weights depends on the estimator implementation. 
In particular, it requires to apply smoothers for the involved outcome regressions (C\ref{cond:smoothers}). 
This excludes methods with non-differentiable objective functions and/or non-linear link functions for outcome prediction, such as Lasso, (penalized) logistic regression, or many neural network architectures. However, it is important to note that the choices for instrument and treatment nuisance parameters do not affect the availability of outcome weights.


Overall, the simple framework of Section \ref{sec:general-framework} proves very handy for compactly deriving new functional forms of outcome weights and recovering known ones. This is interesting and practically useful in its own right, as the obtained weights can be applied in any established weight-based routine.
Additionally, the framework provides a natural lens to investigate basic properties of the outcome weights as we pursue in the following.

\section{Weights properties of pseudo-IV estimators} \label{sec:weight-props-pive}

The results of the previous section enable users to \textit{ex post} inspect whether outcome weights fulfill certain properties. For example, weights adding up to one for the treated (i.e. $\sum_i \omega_i D_i = 1$) and to minus one for the untreated (i.e. $\sum_i \omega_i (1-D_i) = -1$) are often considered desirable because they guarantee certain in- and equivariances of estimators \cite{Imbens2015CausalSciences, Soczynski2024AbadiesEffect}. However, the PIVE framework also allows to analytically investigate the weights properties of estimator implementations. This is conceptually appealing and practically relevant because it permits \textit{ex ante} control over weights properties. Specifically, we investigate under which conditions estimators fulfill one of the five weights properties collected in Table \ref{tab:weights-class} spanned by the total, treated, and untreated weight sums, respectively (see Figure \ref{fig:weights-classes} for a graphical illustration).
\begin{table}[h!]
\centering
        \caption{Outcome weights classification}
    \label{tab:weights-class}
\begin{tabular}{|l|c|c|c|}
\hline
Weights property & $\sum_i \omega_i$  & $\sum_i \omega_i D_i$ & $\sum_i \omega_i (1-D_i)$ \\
\hline
fully-unnormalized & $\neq 0$ & $\neq 1$ & $\neq -1$ \\
untreated-unnormalized & $\neq 0$ & = 1 & $\neq -1$ \\
treated-unnormalized & $\neq 0$ & $\neq 1$ & $= -1$ \\
scale-normalized & = 0 & $= c \neq 1$ & $ = -c \neq -1$ \\
fully-normalized & = 0 & = 1 & $= -1$ \\
\hline
\end{tabular}
\end{table}

The literature documents examples for each class in Table \ref{tab:weights-class}.\footnote{The proposed class labels in Table \ref{tab:weights-class} ensure that all three versions of unnormalized weights would also be labeled as unnormalized by \citeA{Soczynski2024AbadiesEffect}. Although ``normalized'' is a loosely defined term, it seems reasonable in this context to use it for estimators whose outcome weights sum to zero, to remain consistent with previous work.} Fully-unnormalized weights are associated with inverse probability weighting since \citeA{Hajek1971CommentOne}. (Un)treated-unnormalized weights recently appeared in estimators building on Abadie's \citeyear{Abadie2003SemiparametricModels} $\kappa_0$ and $\kappa_1$ where only one group shows weights adding up to (minus) one \cite{Soczynski2024AbadiesEffect}. Scale-normalized weights are described by \citeA{Soczynski2023CovariateEffects} in the context of covariate balancing propensity scores of \citeA{Imai2014CovariateScore}. Such estimators have treated (untreated) weights summing to (minus) the same non-one constant $c$ and also appear prominently in the analysis of partially linear regression based estimators below. Fully-normalized weights are the norm \cite<see overview in>[Ch. 19]{Imbens2015CausalSciences}. 


\subsection{Weights properties in the PIVE framework - general}

The outcome weights properties in Table \ref{tab:weights-class} are determined by three weight sums. This motivates the following protocol to classify PIVE weights:
\begin{enumerate}
    \item Calculate $C := \sum_i \omega_i = \bm{\omega'1_N}$
        \begin{itemize} \singlespacing        
        \item If $C = 0$ $\Rightarrow$ normalized
        \item If $C \neq 0$ $\Rightarrow$ unnormalized
    \end{itemize}
    \item Calculate $C_1 := \sum_i \omega_i D_i = \bm{\omega'D}$
    \begin{itemize} \singlespacing        
        \item If $C \neq 0$ and $C_1 = 1$ $\Rightarrow$ untreated-unnormalized
        \item If $C = 0$ and $C_1 \neq 1$ $\Rightarrow$ scale-normalized
        \item If $C = 0$ and $C_1 = 1$ $\Rightarrow$ fully-normalized
    \end{itemize}
    \item If $C \neq 0$ and $C_1 \neq 1$, calculate $C_0 := \sum_i \omega_i (1-D_i) = \bm{\omega'}(\bm{1_N}-\bm{D})$
        \begin{itemize} \singlespacing
            \item If $C_0 \neq -1$ $\Rightarrow$ fully-unnormalized
            \item If $C_0 = -1$ $\Rightarrow$ treated-unnormalized
        \end{itemize}
\end{enumerate}

Recall from Proposition \ref{prop:iw-pive} that PIVE weights take the form $\bm{\omega'} = (\bm{\tilde{Z}'\tilde{D}})^{-1}\bm{\tilde{Z}'T}$. Therefore classifying the weights properties of an estimator boils down to checking the first two or all of the following equations:
\begin{align}
    (\bm{\tilde{Z}'\tilde{D}})^{-1}\bm{\tilde{Z}'T1_N} & = 0 \label{eq:wp2} \\
    (\bm{\tilde{Z}'\tilde{D}})^{-1}\bm{\tilde{Z}'TD} & = 1 \label{eq:wp1} \\
    (\bm{\tilde{Z}'\tilde{D}})^{-1}\bm{\tilde{Z}'T}(\bm{1_N}-\bm{D}) & = -1 \label{eq:wp3}
\end{align}
This implies that it suffices to investigate the following properties of the transformation matrix as shortcuts to classify the outcome weights:
\begin{enumerate}
    \item $\bm{T1_N} = \bm{0_N}$ because it implies that Equation \ref{eq:wp2} holds \label{item:sc1}
    \item $\bm{TD} = \bm{ \tilde{D}}$ because it implies that Equation \ref{eq:wp1} holds  \label{item:sc2}
    \item $\bm{T}(\bm{1_N}-\bm{D}) = -\bm{\tilde{D}}$ because it implies that Equation \ref{eq:wp3} holds  \label{item:sc3}
\end{enumerate}
This shows how the PIVE structure offers substantial complexity reduction streamlining the derivations below to a large extent. 
It turns out that the weights properties are intimately tied to implementation choices as we first illustrate in the OLS example before moving to more involved cases.

\textit{Example (OLS) continued:}
Only one aspect of the implementation affects OLS weights properties in the sense of Table \ref{tab:weights-class}:
\begin{cond}\label{cond:ols-constant} (covariate matrix with constant) \\
The covariate matrix $\bm{X}$ contains a column of ones, which is by convention the first column. We can therefore write for a matrix with $p$ covariates $\bm{X} (1,\bm{0_p'})' = \bm{1_N}$.
\end{cond}
Condition \ref{cond:ols-constant} is fulfilled in any reasonable application. However, making it explicit illustrates how implementation choices affect weights properties. We start by checking whether the weights sum to zero and use shortcut \ref{item:sc1} focusing on the transformation matrix:
\begin{align*}
    \bm{T}^{ols'} \bm{1_N} &= \bm{M_X} \bm{1_N} =  (\bm{I_N} - \bm{X(X'X)^{-1} X'}) \bm{1_N} \\
    \text{If C\ref{cond:ols-constant}} & = \bm{I_N} \bm{1_N} - \bm{X(X'X)^{-1} X'} \bm{X} (1,\bm{0_p'})' =  \bm{1_N} - \bm{X} (1,\bm{0_p'})' =   \bm{0_N} \Rightarrow \text{normalized}
\end{align*}
We conclude that OLS is always normalized if we include a constant. Next, we investigate whether weights of treated sum up to one via shortcut \ref{item:sc2}:
\begin{align*}
    \bm{T}^{ols'} \bm{D} = \bm{M_X} \bm{D} = \bm{\tilde{D}^{ols}} & \Rightarrow \text{untreated-unnormalized} \\
    \Rightarrow \text{If C\ref{cond:ols-constant}} & \Rightarrow \text{fully-normalized}
\end{align*}
The transformation matrix applied to the treatment recovers the pseudo-treatment, which is sufficient for treated weights adding up to one. Curiously, this holds without further conditions implying that OLS is untreated-unnormalized even without a constant. Both results taken together recover a well-known fact that OLS is fully-normalized under Condition \ref{cond:ols-constant}. Additionally, following the proposed framework step by step uncovers a nuisance regarding the case without a constant.

\subsection{Weights properties in the PIVE framework - concrete} \label{sec:ow-props-concrete}

\subsubsection{Implementation details} \label{sec:imp-detail}

This section collectively introduces implementation details that become relevant in the later derivations. We start with a relatively mild condition:\begin{cond}\label{cond:affine-smooth} (affine smoother matrix) \\
In addition to Condition \ref{cond:smoothers}, all rows of the smoother matrices add up to one: 
$$\bm{S 1_N} = \bm{S_0^d 1_N} = \bm{S_1^d 1_N} = \bm{S_0^z 1_N} = \bm{S_1^z 1_N} = \bm{1_N}$$
\end{cond}
Most smoothers discussed in the literature fulfill this property. However, \citeA{Curth2024WhySmoothers} note that boosted trees can be an exception.

The next condition is relevant for the treatment group specific outcome nuisances:
\begin{cond}\label{cond:nsbg} (no smoothing between treatment groups) \\
In addition to Condition \ref{cond:smoothers}b, the treatment group specific predictions are formed using only observations of the respective group.  This ensures that
\begin{align}
\bm{S^d_1 1_N} = \bm{S^d_1D} \Rightarrow \bm{S^d_1 (1_N - D)} = \bm{0_N} \\
\bm{S^d_0 1_N} = \bm{S^d_0(1_N-D)} \Rightarrow \bm{S^d_0 D} = \bm{0_N}
\end{align}
\end{cond}
This condition is relevant for AIPW estimators and in line with standard implementations forming the group specific outcome models in the respective subgroups. 

The next condition is less familiar but important for estimators based on partially linear regression and Wald-AIPW:
\begin{cond}\label{cond:osmatt} (outcome smoother matrix applied to treatment) \\
(C5a) The treatment predictions are formed using the outcome smoother matrix:
\begin{equation}
\bm{SD} = \bm{\hat{D}}
\end{equation}
(C5b) The treatment predictions in the different instrument groups are formed using the respective outcome smoother matrix:
\begin{equation}
\bm{S^z_1D} = \bm{\hat{D}^z_1} \text{~and~} \bm{S^z_0D} = \bm{\hat{D}^z_0}
\end{equation}
\end{cond}
This goes against the idea of many flexible estimators to entertain different models for outcome and treatment predictions, respectively. Therefore, this condition is not in line with standard implementations.

The final condition is relevant for all estimators involving an inverse probability weighting component:
\begin{cond}\label{cond:norm-ipw} (normalized inverse probability weights) \\
(C6a) $\lambda_{0,i}^{norm} :=  \lambda_{0,i}^{ipw} / \E_N[\lambda_{0,i}^{ipw}]$ and $\lambda_{1,i}^{norm} :=  \lambda_{1,i}^{ipw} / \E_N[\lambda_{1,i}^{ipw}] \Rightarrow \bm{1_N'\lambda_{1}^{norm}} = \bm{1_N' \lambda_{0}^{norm}} = N$ \\
(C6b) $\lambda_{0,i}^{norm,z} :=  \lambda_{0,i}^{ipw,z} / \E_N[\lambda_{0,i}^{ipw,z}]$ and $\lambda_{1,i}^{norm,z} :=  \lambda_{1,i}^{ipw,z} / \E_N[\lambda_{1,i}^{ipw,z}] \Rightarrow \bm{1_N'\lambda_{1}^{norm,z}} = \bm{1_N'\lambda_{0}^{norm,z}} = N$
\end{cond}
C6a is the standard \citeA{Hajek1971CommentOne} normalization and usually recommended in applications \cite{busso2014new}. C6b is suggested by \citeA{Uysal2011ThreeMethods} and recommended by \citeA{Soczynski2024AbadiesEffect}.


\subsubsection{Weights properties of Instrumental Forest and its special cases} \label{sec:iw-probs-if}

Without further conditions $\bm{\omega}^{if}(\bm{x})$ in \eqref{eq:if} is fully-unnormalized. In the following, we explore conditions leading to (fully-)normalized weights. First, we investigate how $\bm{T^{if} 1_N} = \bm{0_N}$ could be obtained:
\begin{align*}
    \bm{T^{if} 1_N} &= (\bm{I_N} - \bm{S}) \bm{1_N} = \bm{1_N} - \bm{S} \bm{1_N}  \\
    \text{If C\ref{cond:affine-smooth}} &= \bm{1_N} - \bm{1_N} = \bm{0_N} \Rightarrow \text{normalized}
\end{align*}
This establishes that the standard implementation of IF in \texttt{grf} uses normalized weights because it applies the affine smoother random forest (C\ref{cond:affine-smooth}) to estimate the outcome nuisance.

The next question is when treated weights sum to one. To this end, it is sufficient to understand when $\bm{T^{if} D} = \bm{\tilde{D}^{if}}$:
\vspace{-.5cm}
\begin{align*}
    \bm{T^{if} D} &= (\bm{I_N} - \bm{S}) \bm{D} =  \bm{D} - \bm{SD} & \\
    \text{If C\ref{cond:osmatt}a} &= \bm{D} - \bm{\hat{D}} = \bm{\hat{V}}  = \bm{\tilde{D}^{if}} \Rightarrow \text{untreated-unnormalized} \\
     & \Rightarrow \text{ If C\ref{cond:affine-smooth} \& C\ref{cond:osmatt}a} \Rightarrow \text{fully-normalized}
\end{align*}
The two results span different scenarios. The practically relevant one being that Condition \ref{cond:affine-smooth} holds but Condition \ref{cond:osmatt}a does not because different treatment and outcome models are applied. This means that in practice IF weights are only scale-normalized but not fully-normalized. Only applying the same affine smoother matrix to predict outcome and treatment ensures fully-normalized weights.\footnote{Curiously, applying the same non-affine smoother ensures that at least treated weights sum to one generalizing the observation regarding OLS without constant in the previous section.}

Recall from Table \ref{tab:if-paths} that CF, PLR-IV and PLR use the same transformation matrix as IF. Consequently, they are also scale-normalized in standard applications. In contrast, OLS and TSLS apply the same projection matrix to form treatment and outcome predictions such that C\ref{cond:osmatt}a holds by construction. Again the observations regarding OLS in the previous section immediately apply for TSLS because they share pseudo-treatment and transformation matrix. Also TSLS with a constant is fully-normalized and untreated-unnormalized without a constant. 

For completeness observe that the difference in means estimator fulfils by construction both Conditions \ref{cond:affine-smooth} and \ref{cond:osmatt}a, and is therefore always fully-normalized. An overview of conditions and weights properties is collected in Table \ref{tab:iw-props} below.

\subsubsection{Weights properties of AIPW and its special cases} \label{sec:aipw-prop}

First, we investigate under which conditions AIPW is normalized:
\begin{align*}
    \bm{T^{aipw} 1_N} 
    &= \bm{S^d_1}\bm{1_N} - \bm{S^d_0}\bm{1_N} + \bm{\lambda_{1}^{ipw}} - diag(\bm{\lambda_{1}^{ipw}}) \bm{S^d_1} \bm{1_N} -  \bm{\lambda_{0}^{ipw}} + diag(\bm{\lambda_{0}^{ipw}}) \bm{S^d_0} \bm{1_N} \\
    \text{If C\ref{cond:affine-smooth}} & = \bm{1_N} - \bm{1_N} + \bm{\lambda_{1}^{ipw}} - \bm{\lambda_{1}^{ipw}} -  \bm{\lambda_{0}^{ipw}} + \bm{\lambda_{0}^{ipw}} = \bm{0_N} \Rightarrow \text{normalized}
\end{align*}
This result contains two surprising components. First, we did \textit{not} apply normalized IPW weights (C\ref{cond:norm-ipw}a) to achieve normalized AIPW. This means AIPW is self-normalizing once affine smoothers are applied. Second, normalized IPW weights alone do not normalize AIPW weights as a similar simplification is not possible under C\ref{cond:norm-ipw}a only.

The second step investigates when treated weights sum to one:
\begin{align*}
    \bm{T^{aipw} D} 
    &= \bm{S^d_1} \bm{D} - \bm{S^d_0} \bm{D} + \bm{\lambda_{1}^{ipw}} - diag(\bm{\lambda_{1}^{ipw}}) \bm{S^d_1}  \bm{D} + diag(\bm{\lambda_{0}^{ipw}}) \bm{S^d_0} \bm{D} \\
    \text{If C\ref{cond:nsbg}} & = \bm{S^d_1} \bm{1_N} + \bm{\lambda_{1}^{ipw}} - diag(\bm{\lambda_{1}^{ipw}}) \bm{S^d_1}  \bm{\bm{1_N}} \\
    \text{If C\ref{cond:affine-smooth} \& C\ref{cond:nsbg}} & = \bm{1_N} + \bm{\lambda_{1}^{ipw}} - \bm{\lambda_{1}^{ipw}} = \bm{1_N} = \bm{\tilde{D}^{aipw}} \Rightarrow \text{fully-normalized }
\end{align*}
This means that standard implementations using affine smoothers to estimate outcome nuisances in the (un)treated groups separately are self-fully-normalizing regardless which IPW weights are applied. This implies that RA inherits weights properties from AIPW because it can be considered as applying IPW weights of zero (see Table \ref{tab:aipw-paths}). In contrast, IPW can be considered as applying smoother matrices of zeros. These uninformative smoother matrices by construction fulfill C\ref{cond:nsbg} but not C\ref{cond:affine-smooth} such that IPW weights are not (fully-)normalized. This recovers the well-known result of \citeA{Hajek1971CommentOne} regarding IPW as a special case of AIPW. Obviously IPW with explicitly fully-normalized weights (under C\ref{cond:norm-ipw}a) are fully-normalized.\footnote{To see this within the framework note that under C\ref{cond:norm-ipw}a $\bm{1_N' diag(\bm{\lambda_{1}^{norm}} - \bm{\lambda_{0}^{norm}}) 1_N} = N - N = 0$ and $\bm{diag(\bm{\lambda_{1}^{norm}} - \bm{\lambda_{0}^{norm}}) D} = \bm{1_N} = \bm{\tilde{D}^{aipw}}$ establishing full-normalization.}


\subsubsection{Weights properties of Wald-AIPW and its special cases} \label{sec:wald-aipw-prop}

We can not directly apply the results of Section \ref{sec:aipw-prop} because the pseudo-outcome and -treatment differ. However, to show that the estimator is normalized if affine smoothers are applied for the outcome regressions requires only to change the superscripts:
\begin{align*}
    \bm{T^{iv-aipw} 1_N} 
    &= \bm{S^z_1}\bm{1_N} - \bm{S^z_0}\bm{1_N} + \bm{\lambda_{1}^{ipw,z}} - diag(\bm{\lambda_{1}^{ipw,z}}) \bm{S^z_1} \bm{1_N} -  \bm{\lambda_{0}^{ipw,z}} + diag(\bm{\lambda_{0}^{ipw,z}}) \bm{S^z_0} \bm{1_N} \\
    \text{If C\ref{cond:affine-smooth}} & = \bm{1_N} - \bm{1_N} + \bm{\lambda_{1}^{ipw,z}} - \bm{\lambda_{1}^{ipw,z}} -  \bm{\lambda_{0}^{ipw,z}} + \bm{\lambda_{0}^{ipw,z}} = \bm{0_N} \Rightarrow \text{normalized}
\end{align*}
However, the investigation of the sum of treated weights shows notable differences:
\begin{align*}
    \bm{T^{iv-aipw} D} 
    &=  \bm{S^z_1} \bm{D} - \bm{S^z_0} \bm{D} + diag(\bm{\lambda_{1}^{ipw,z}}) (\bm{D} - \bm{S^z_1}\bm{D}) -  diag(\bm{\lambda_{0}^{ipw,z}}) (\bm{D} - \bm{S^z_0} \bm{D})] \\
    \text{If C\ref{cond:osmatt}b} & 
    =  \bm{\hat{D}^z_1} - \bm{\hat{D}^z_0} + diag(\bm{\lambda_{1}^{ipw,z}}) (\bm{D} - \bm{\hat{D}^z_1}) -  diag(\bm{\lambda_{0}^{ipw,z}}) (\bm{D} - \bm{\hat{D}^z_0})] \\
    & = \bm{\tilde{D}^{iv-aipw}} \Rightarrow \text{untreated-unnormalized } \\
    & \Rightarrow \text{If C\ref{cond:affine-smooth} \& C\ref{cond:osmatt}b} \Rightarrow \text{fully-normalized }
\end{align*}
Wald-AIPW is therefore only scale-normalized unless we apply the outcome smoothers to also predict the treatments. This goes against the idea of using different models for each nuisance parameter. Unlike AIPW, Wald-AIPW is therefore not expected to be fully-normalized in standard applications. Another point worth noting is that separating the sample by instrument value when estimating outcome/treatment nuisances - an IV version of C\ref{cond:nsbg} - is not sufficient to achieve fully-normalized weights of Wald-AIPW.

Similar to the previous section Wald-RA inherits all properties from Wald-AIPW. However, Wald-IPW is always untreated-unnormalized because it can be considered as applying the same zero smoother matrix to outcome and treatment (C\ref{cond:osmatt}b). Additionally normalizing the weights (C\ref{cond:norm-ipw}b) makes Wald-IPW even fully-normalized.\footnote{This follows by considering the full numerator of the weight and not only the transformation matrix such that $\bm{1_N' diag(\bm{\lambda_{1}^{norm,z}} - \bm{\lambda_{0}^{norm,z}}) 1_N} = N - N = 0$ due to Condition \ref{cond:norm-ipw}b.} This recovers observations regarding Wald-IPW by \citeA{Soczynski2024AbadiesEffect} within the framework of this paper. Also their findings regarding Abadie's \citeyear{Abadie2003SemiparametricModels} $\kappa$ estimators can be obtained in the framework as shown in Appendix \ref{app:replicate-usw}.

\subsection{Consolidation}

Table \ref{tab:iw-props} summarizes the sufficient conditions for closed-form and (fully-)normalized outcome weights.\footnote{Table \ref{tab:app-iw-props} in the Appendix provides an extended table collecting which conditions are fulfilled by construction and including results for (un)treated-unnormalized weights for completeness. However, those are rather of academic value and we focus on the practically relevant cases in the main text.} Estimators with a check mark in the second column always have a weighted representation. Those are the estimators based on IPW and OLS where the weights are either obvious or at least well-studied \cite<e.g.>{Imbens1997EstimatingModels,Imbens2015CausalSciences, Imbens2015MatchingExamples, Chattopadhyay2023OnInference}.
They are still included to demonstrate the generality of the framework but not to provide new insights.
Those are obtained for more sophisticated outcome adaptive estimators for which weighted representations are not available in the literature. 

\begin{table}
\centering
\begin{threeparttable} 
\onehalfspacing
\caption{Conditions for closed-form and properties of outcome weights} \label{tab:iw-props}
\begin{tabular}{lccc}   
\toprule
\textbf{Estimator} & \textbf{Closed-form} & \textbf{Normalized} & \textbf{Fully-normalized}  \\ 
\midrule
Instrumental forest & C\ref{cond:smoothers}a & C\ref{cond:affine-smooth} & C\ref{cond:affine-smooth} \& C\ref{cond:osmatt}a  \\ 
PLR-IV & C\ref{cond:smoothers}a & C\ref{cond:affine-smooth} & C\ref{cond:affine-smooth} \& C\ref{cond:osmatt}a  \\ 
TSLS & \cmark & C\ref{cond:ols-constant} & C\ref{cond:ols-constant} \\ 
Wald & \cmark & \cmark & \cmark \\ 
Causal Forest & C\ref{cond:smoothers}a & C\ref{cond:affine-smooth} & C\ref{cond:affine-smooth} \& C\ref{cond:osmatt}a  \\
PLR & C\ref{cond:smoothers}a & C\ref{cond:affine-smooth} & C\ref{cond:affine-smooth} \& C\ref{cond:osmatt}a  \\
OLS & \cmark & C\ref{cond:ols-constant} & C\ref{cond:ols-constant} \\ 
DiM & \cmark & \cmark & \cmark \\ 
AIPW & C\ref{cond:smoothers}b & C\ref{cond:affine-smooth} & C\ref{cond:affine-smooth} \& C\ref{cond:nsbg}  \\
RA & C\ref{cond:smoothers}b & C\ref{cond:affine-smooth} & C\ref{cond:affine-smooth} \& C\ref{cond:nsbg}  \\ 
IPW & \cmark & C\ref{cond:norm-ipw}a & C\ref{cond:norm-ipw}a \\ 
Wald-AIPW & C\ref{cond:smoothers}c & C\ref{cond:affine-smooth} & C\ref{cond:affine-smooth} \& C\ref{cond:osmatt}b  \\
Wald-RA & C\ref{cond:smoothers}c & C\ref{cond:affine-smooth} & C\ref{cond:affine-smooth} \& C\ref{cond:osmatt}b  \\
Wald-IPW & \cmark & C\ref{cond:norm-ipw}b & C\ref{cond:norm-ipw}b \\ 
\bottomrule
\end{tabular}
\begin{tablenotes} \small \item \textit{Abbreviations:} DiM: difference in means; RA: regression adjustment; IPW: inverse probability weighting; AIPW: augmented IPW; OLS: ordinary least squares; PLR: partially linear regression; TSLS: two-stage least squares \end{tablenotes}  
\end{threeparttable}    
\end{table}

The results collected in Table \ref{tab:iw-props} highlight the crucial role of implementation details for availability and properties of outcome weights. First, column two documents that researchers can ensure that outcome weights can be accessed \textit{ex post} by applying smoothers to form outcome predictions as shown in Section \ref{sec:iw}. Second, estimator specific implementation decisions \textit{ex ante} determine certain weights properties. Columns three and four of Table \ref{tab:iw-props} can serve as look-up table for researchers who want to ensure that a particular implementation of an estimator generates outcome weights of a desired class. They contain several surprising or at least undocumented results regarding six prominent DML and GRF instances:
\begin{enumerate}
    \item PLR(-IV), causal/instrumental forests, and Wald-AIPW are not fully-normalized in standard implementations because they usually apply different treatment and outcome models (C\ref{cond:osmatt} not fulfilled).
    \item AIPW is fully-normalized in standard implementations because they usually apply affine smoothers and estimate treated and untreated outcomes separately (C\ref{cond:affine-smooth} \& C\ref{cond:nsbg} fulfilled).
\end{enumerate}

\subsection{Empirical Monte Carlo illustration} \label{sec:emcs}

This section runs an Empirical Monte Carlo Study (EMCS) to illustrate that most standard implementations of DML and GRF are not fully-normalized. EMCS take a real dataset and modify some components such that the ground truth is known in the semi-synthetic dataset \cite<e.g.>{Huber2013,Wendling2018}. Here, we use the treatment, instrument, and covariates of the 401(k) data \cite{Chernozhukov2016High-DimensionalR} but with a noiseless outcome $Y_i^* = 1 + D_i$. This simulates the most powerful treatment leaving every untreated unit at one and shifting every treated unit to two. We expect estimators to estimate an effect of exactly one in this setting without outcome noise.  However, only fully-normalized implementations are guaranteed to achieve this because for them, $\bm{\omega' Y^*} = \bm{\omega'} (\bm{1_N} + \bm{D}) = 1$. 

This exercise is run with the \texttt{DoubleML} \cite{Bach2024DoubleML:R} and the \texttt{grf} \cite{Tibshirani2024Grf:Forests} R packages applied to 100 bootstrap samples. The nuisance parameters in \texttt{DoubleML} are obtained using honest random forest (affine smoother) or XGBoost (non-affine smoother). Each function uses its default values. Table \ref{tab:emcs-estimators} summarizes the ten implementations under consideration. The final column shows whether an implementation is fully-normalized according to the theoretical results in Table \ref{tab:iw-props} and therefore expected to find the ``effect'' of one.

\begin{table}
\begin{threeparttable}
\onehalfspacing \small
\caption{EMCS estimators and their labels} \label{tab:emcs-estimators}
\begin{tabular}{lcccc} 
\toprule
\textbf{Label} & \textbf{Estimator} & \textbf{Package} & \textbf{Nuisance} & \textbf{Fully-normalized?}\\ 
\midrule
PLR DML RF & PLR & \texttt{DoubleML} & random forest & no b/c \cancel{C\ref{cond:osmatt}a} \\ 
PLR DML XGB & PLR & \texttt{DoubleML} & XGBoost & no b/c \cancel{C\ref{cond:affine-smooth}} \& \cancel{C\ref{cond:osmatt}a} \\ 
AIPW DML RF & AIPW & \texttt{DoubleML} & random forest & yes \\ 
AIPW DML XGB & AIPW & \texttt{DoubleML} & XGBoost & no b/c \cancel{C\ref{cond:affine-smooth}} \\ 
CF grf RF & CF & \texttt{grf} & random forest & no b/c \cancel{C\ref{cond:osmatt}a} \\ 
PLR-IV DML RF & PLR-IV & \texttt{DoubleML} & random forest & no b/c \cancel{C\ref{cond:osmatt}a} \\ 
PLR-IV DML XGB & PLR-IV & \texttt{DoubleML} & XGBoost & no b/c \cancel{C\ref{cond:affine-smooth}} \& \cancel{C\ref{cond:osmatt}a} \\ 
Wald-AIPW DML RF & Wald-AIPW & \texttt{DoubleML} & random forest & no b/c \cancel{C\ref{cond:osmatt}b} \\ 
Wald-AIPW DML XGB & Wald-AIPW & \texttt{DoubleML} & XGBoost & no b/c \cancel{C\ref{cond:affine-smooth}} \& \cancel{C\ref{cond:osmatt}b} \\ 
IF grf RF & IF & \texttt{grf} & random forest & no b/c \cancel{C\ref{cond:osmatt}a} \\ 
\bottomrule
\end{tabular}
\begin{tablenotes}
\item \textit{Notes:} The columns show (i) the labels used in Figure \ref{fig:illu-401k}, (ii) which estimator defined in Section \ref{sec:ow-concrete} is applied, (iii) the applied R package, (iv) the nuisance parameters, and (v) why the specific implementation are (not) expected to be fully-normalized.
\end{tablenotes}
\end{threeparttable}
\end{table}

\begin{figure}
    \centering        
    \begin{minipage}{\linewidth} 
    \caption{Empirical Monte Carlo illustration}
        \label{fig:illu-401k}
        \includegraphics[width=\linewidth]{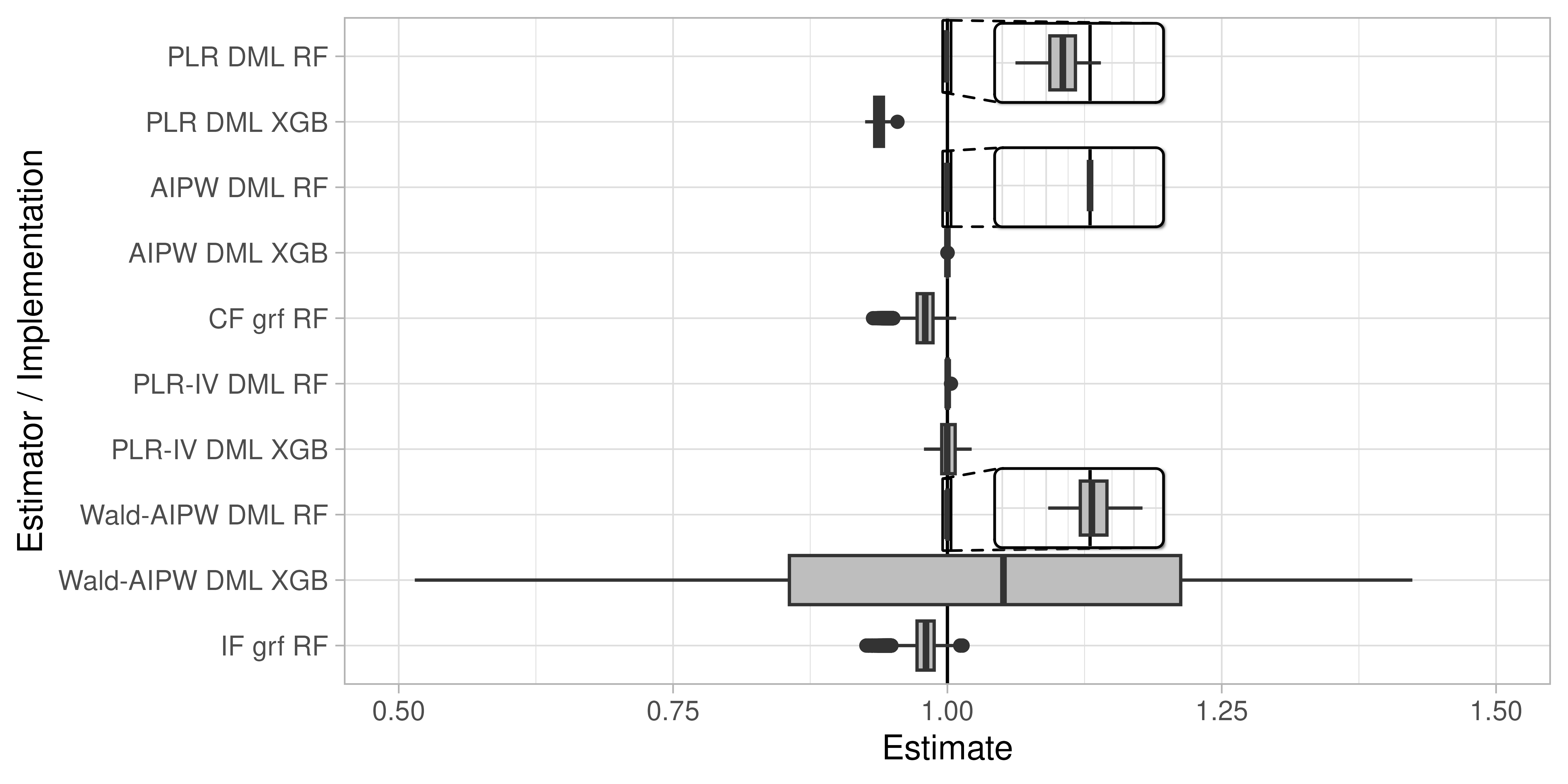}       
        \small 
        \textit{Notes:} Boxplots show the results of 100 bootstraps of the 401(k) data \cite{Chernozhukov2016High-DimensionalR} where the outcome is set to $Y_i^* = 1 + D_i$. The estimators are implemented using the default settings of the \texttt{DoubleML} and  \texttt{grf} packages (see Table \ref{tab:emcs-estimators} for the labels). The causal/instrumental forest produces 9,915 estimates per replication such that their boxplots are based on $\sim1$ million estimates. The simulated effect is always one indicated by the solid line. The shadowed boxes in rows 1, 3 and 9 zoom into the range between 0.996 and 1.003. 49 outliers of Wald-AIPW DML XGB ranging from -16 to 55 are omitted. See \href{https://mcknaus.github.io/assets/code/Notebook_EMCS_illustration_401k.nb.html}{EMCS R notebook} for the code and more details.
    \end{minipage}
\end{figure}

The theoretical predictions are confirmed in Figure \ref{fig:illu-401k}. The boxplots show that only AIPW with an affine smoother finds an effect of exactly one in all bootstrap samples. The other methods deviate from one to varying degrees. The XGBoost Wald-AIPW stands out in estimating effects between -16 and 55 (the graph is truncated). However, also causal/instrumental forests estimate heterogeneous effects between 0.93 and 1.01 although there is no heterogeneity to be found in the provided data. This illustrates the theoretical findings even for \texttt{DoubleML} implementations where the extraction of the outcome weights is currently not possible because the required smoother matrices are not accessible.

\section{Application: 401(k) covariate balancing} \label{sec:application}

The novel outcome weights for DML and GRF can be used in established routines or to develop estimator-specific applications. We illustrate the former with covariate balancing, leaving the latter for future research. As in Section \ref{sec:emcs}, we use the 401(k) data from \citeA{Chernozhukov2016High-DimensionalR}, but this time with the real outcome ``net assets''. 

\subsection{Average effects}

\begin{figure}[ht]
    \centering        
    \begin{minipage}{\linewidth} 
    \caption{Covariate balancing plots - average effects}
        \label{fig:balance-401k}
        \includegraphics[width=\linewidth]{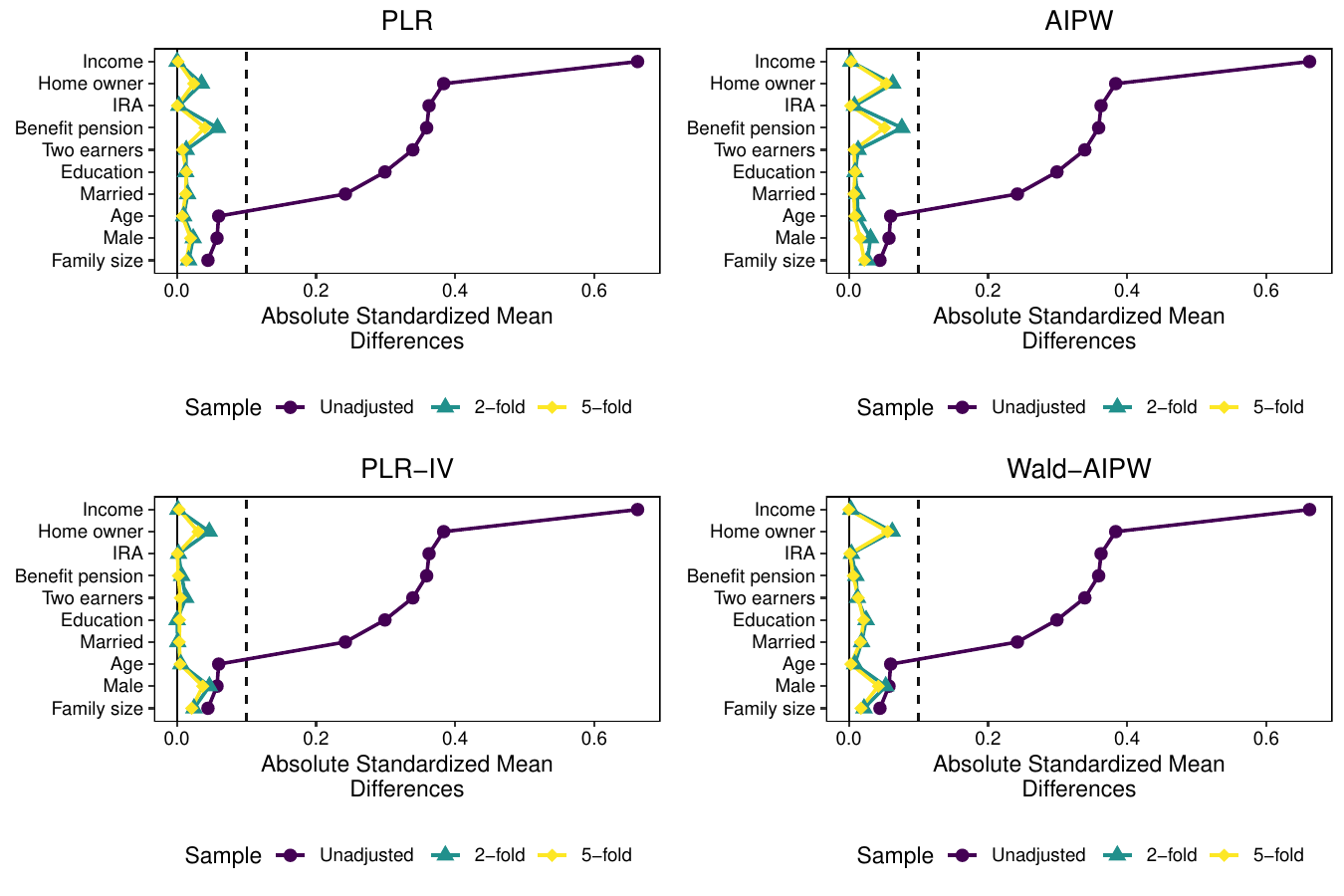}       
        \small 
        \textit{Notes:} Each plot is created with the \texttt{love.plot()} function of the \texttt{cobalt} R package \cite{Greifer2024Cobalt:Plots} and the weights derived in Section \ref{sec:ow-concrete}.
    \end{minipage}
\end{figure}

First, we investigate covariate balancing for DML estimated average effects. PLR(-IV) and (Wald-)AIPW are implemented using honest random forests with 2- and 5-fold cross-fitting.
Figure \ref{fig:balance-401k} presents canonical balancing plots from the \texttt{cobalt} R package \cite{Greifer2024Cobalt:Plots} displaying absolute standardized mean differences (SMD).
We observe that each method successfully balances the previously unbalanced covariates, in particular the income variable. Furthermore, cross-fitting with 5-folds achieves better covariate balancing compared to 2-folds. This demonstrates how DML outcome weights can be utilized in the design phase described by \citeA{Rubin2007TheTrials}, allowing researchers to commit to the preferred implementation before examining the results.

The supplementary \href{https://mcknaus.github.io/assets/code/Notebook_Application_average_401k.nb.html}{average effects R notebook} also provides point estimates and additional results, such as showing that 10 cross-fitting folds provide no further improvement over 5 folds and that the scale-normalized weights sum to values close to one (0.995 and closer).

\subsection{Causal forest}

\begin{figure}
    \centering        
    \begin{minipage}{\linewidth} 
    \caption{Covariate balancing plots - heterogeneous effects}
        \label{fig:balance-cate-401k}
        \includegraphics[width=\linewidth]{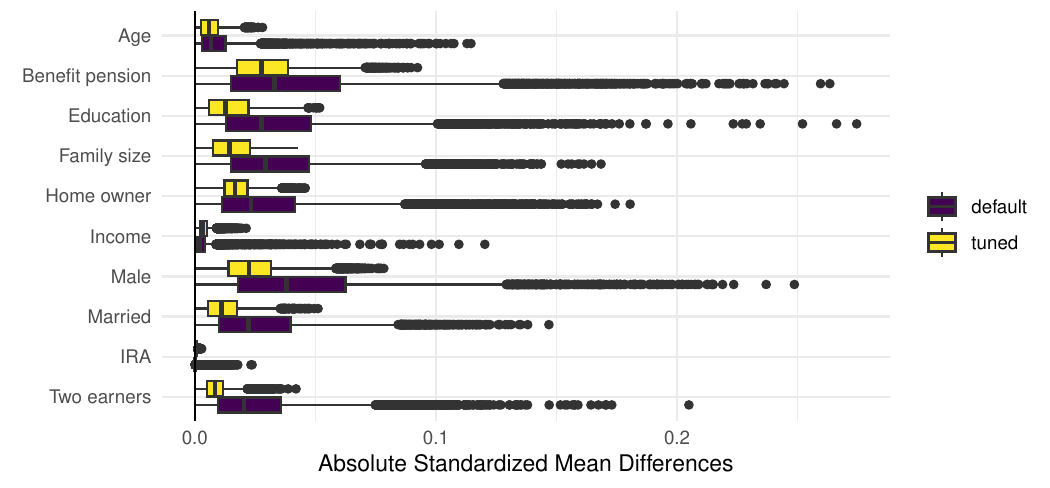}       
        \small 
        \textit{Notes:} Boxplots of absolute standardized mean differences for conditional average treatment effects estimated by causal forest using the default and tuned hyperparameters.
    \end{minipage}
\end{figure}

Checks like those in Figure \ref{fig:balance-401k} are standard when estimating average effects. Similarly, we can assess covariate balancing for all 9,915 conditional average treatment effects (CATEs) produced by the \texttt{causal\_forest()} function of the \texttt{grf} package. As an illustration, we investigate the importance of hyperparameter tuning for causal forests by comparing the default implementation with \texttt{tune.parameters = "all"}. Figure \ref{fig:balance-cate-401k} shows boxplots of absolute standardized mean differences (SMDs) for each CATE estimate. The results highlight that tuning the forests substantially improves covariate balancing in this application. The tuned version achieves absolute SMDs of 0.1 or lower, whereas the default settings frequently exceed this threshold, with some values even above 0.2. This highlights how standard diagnostics for average effects can also be applied to CATE estimates.

The supplementary \href{https://mcknaus.github.io/assets/code/Notebook_Application_heterogeneous_401k.nb.html}{heterogeneous effects R notebook} reveals that the imbalances in the default forest coincide with implausible effect sizes ranging from -\$21k to \$78k, whereas the tuned forest yields more plausible estimates between \$8k and \$23k. This highlights the importance of parameter tuning for causal forests in this application. A similar pattern is observed for the instrumental forest, though with higher levels of |SMD|.

The supplementary notebook additionally examines descriptive statistics of the outcome weights multiplied by $2D_i-1$ to switch the sign of the untreated weights for better comparability. It documents that (i) both causal forests use negative weights, though to a limited extent, (ii) instrumental forests assign substantial negative weights to never-takers, consistent with the outcome weights in \citeA{Imbens1997EstimatingModels} and the fact that the 401(k) setting has no always-takers by design, (iii) tuned forests use much smaller weights in absolute values, indicating more stable and reliable estimates, (iv) the sum of weights ranges from 0.98 to 1.02 for the default settings and from 0.995 to 1.005 for the tuned forest, making the tuned forest approximately fully-normalized in this application. Future work should explore whether this represents a general pattern.

\section{Closing remarks} \label{sec:conc}

More estimators than previously noted can be expressed as weighted outcomes. The paper provides a general framework and derives novel weights for double machine learning and generalized random forest estimators. A key learning is that both availability and properties of the outcome weights depend on implementation choices and are therefore controlled by the user.

The paper focuses on providing general theoretical tools and standard illustrations. This acknowledges that access to their closed-form expressions is a prerequisite for developing new use cases or theoretical results for outcome weights. With the provided tools now available, many follow-up questions arise for future research:
\begin{itemize}
    \item Are there estimator specific use cases beyond the standard diagnostic tools?
    \item What are the closed-form expressions and properties of other PIVE outcome weights?
    \item Does the finding that several popular estimators do not use fully-normalized weights challenge the preference for such weights in the literature, or could explicitly normalizing the weights improve the finite sample performance of these estimators?
    \item Does the need to restrict outcome predictors to smoothers for access to outcome weights suggest a trade-off between interpretability and performance for outcome adaptive causal effect estimators?
    \item Do the provided outcome weights have implications for statistical inference, asymptotic properties, or double robustness robustness properties?
\end{itemize}
The investigation of the latter point most likely requires to restrict focus to analytically tractable smoothers in contrast to the generic smoothers allowed for in this paper. The fact that the smoothers and therefore the outcome weights may depend on the outcome pose non-trivial challenges. For example, it makes the outcome weights not compatible with approaches to use them for statistical inference as the existing approaches require outcome weights to be independent of the outcomes \cite<e.g.>[Ch. 19]{Imbens2015CausalSciences}. Tailored sample splits as in \citeA{Lechner2018} could ensure the required independence but explorations along these lines are left for future work.

\newpage
\onehalfspacing
\bibliographystyle{apacite}
\renewcommand{\APACrefYearMonthDay}[3]{\APACrefYear{#1}}
\bibliography{references}

\newpage
\begin{appendices}
\doublespacing
\numberwithin{equation}{section}
\counterwithin{figure}{section}
\counterwithin{table}{section}

\section{Supplementary Appendix}

\subsection{Estimators under consideration} \label{sec:app-estimators}

\subsubsection{Motivating target parameters}

Let $Y_i(1)$ and $Y_i(0)$ be the potential outcomes under treatment and control, respectively. The paper is motivated by estimators of causal effects that aggregate the individual treatment effects $Y_i(1) - Y_i(0)$ over different populations:
\begin{itemize}
\item $\E[Y_i(1) - Y_i(0)]$, the average treatment effect (ATE)
\item $\E[Y_i(1) - Y_i(0) \mid \bm{X_i} = \bm{x}]$, the conditional ATE (CATE)
\item $\E[Y_i(1) - Y_i(0) \mid Complier_i]$, the local ATE (LATE), where $Complier_i$ is the subgroup being shifted into treatment by a binary instrument \cite{Angrist1996IdentificationVariables}
\item $\E[Y_i(1) - Y_i(0) \mid Complier_i, \bm{X_i} = \bm{x}]$, the conditional local ATE (CLATE)
\end{itemize}
Alternatively, we might impose a partially linear outcome model $Y_i = \theta D_i + g(\bm{X_i}) + \epsilon_i$ and aim to estimate $\theta$. 

Definition and identification of such parameters are discussed in detail in the literature and in textbooks. However, the numerical results provided in the main text also apply if the identifying assumptions do not hold, the target is explicitly non-causal, or the target is a different causal quantity.  

\subsubsection{Estimators}

Table \ref{tab:estimators} collects how the considered estimators differ in the aggregation level of the target effect (average or conditional effects), the research design in which they are usually applied (randomized controlled trials, unconfoundedness or instrumental variables), and regarding outcome modelling assumptions (none, partially linear, or linear models). 
\begin{center}
\begin{landscape}
\begin{threeparttable}[ht] 
\onehalfspacing
\caption{Overview of considered estimators} \label{tab:estimators}
\begin{tabular}{lcccc}   
\toprule
& \textbf{Aggregation} & \textbf{Research} & \textbf{Outcome} & \textbf{Outcome weights in the literature} \\ 
\textbf{Estimator}       & \textbf{level}       & \textbf{design}    & \textbf{model} &  \\ 
\midrule
DiM                      & Average              & RCT                & none            & Imbens \& Rubin \citeyear{Imbens2015CausalSciences}, Ch. 19.4 \\ 
RA                       & Average              & RCT/UC             & none            & - \\ 
IPW                      & Average              & RCT/UC             & none            & Horvitz \& Thompson \citeyear{Horvitz1952} \\ 
AIPW                     & Average              & RCT/UC             & none            & \makecell{Knaus \citeyear{Knaus2021ASkills} (Post-Lasso) \\ Chattopadhyay \& Zubizarreta \citeyear{Chattopadhyay2023OnInference} (OLS)} \\ 
PLR                      & Average              & RCT/UC             & partially linear& - \\ 
OLS                      & Average              & RCT/UC             & linear          & Chattopadhyay \& Zubizarreta \citeyear{Chattopadhyay2023OnInference}  \\ 
Wald                     & Average              & IV                 & none            & Imbens \& Rubin \citeyear{Imbens1997EstimatingModels} \\ 
Wald-RA                  & Average              & IV                 & none            & - \\ 
Wald-IPW                 & Average              & IV                 & none            & Abadie \citeyear{Abadie2003SemiparametricModels} \\ 
Wald-AIPW                & Average              & IV                 & none            & - \\ 
PLR-IV                   & Average              & IV                 & partially linear& - \\ 
TSLS                     & Average              & IV                 & linear          & Chattopadhyay \& Zubizarreta \citeyear{Chattopadhyay2021OnInference} \\ 
Causal Forest            & Conditional          & RCT/UC             & none            & - \\ 
Instrumental Forest      & Conditional          & IV                 & none            & - \\ 
\bottomrule
\end{tabular}
\begin{tablenotes} \item \textit{Abbreviations:} DiM: difference in means; RA: regression adjustment; IPW: inverse probability weighting; AIPW: augmented IPW; OLS: ordinary least squares; PLR: partially linear regression; TSLS: two-stage least squares; RCT: randomized controlled trials; UC: unconfoundedness; IV: instrumental variable \end{tablenotes}  
\end{threeparttable}
\end{landscape}
\end{center}

\begin{figure}
    \caption{Considered estimators and connections between them}
        \label{fig:estimators}
        \includegraphics[width=1\linewidth]{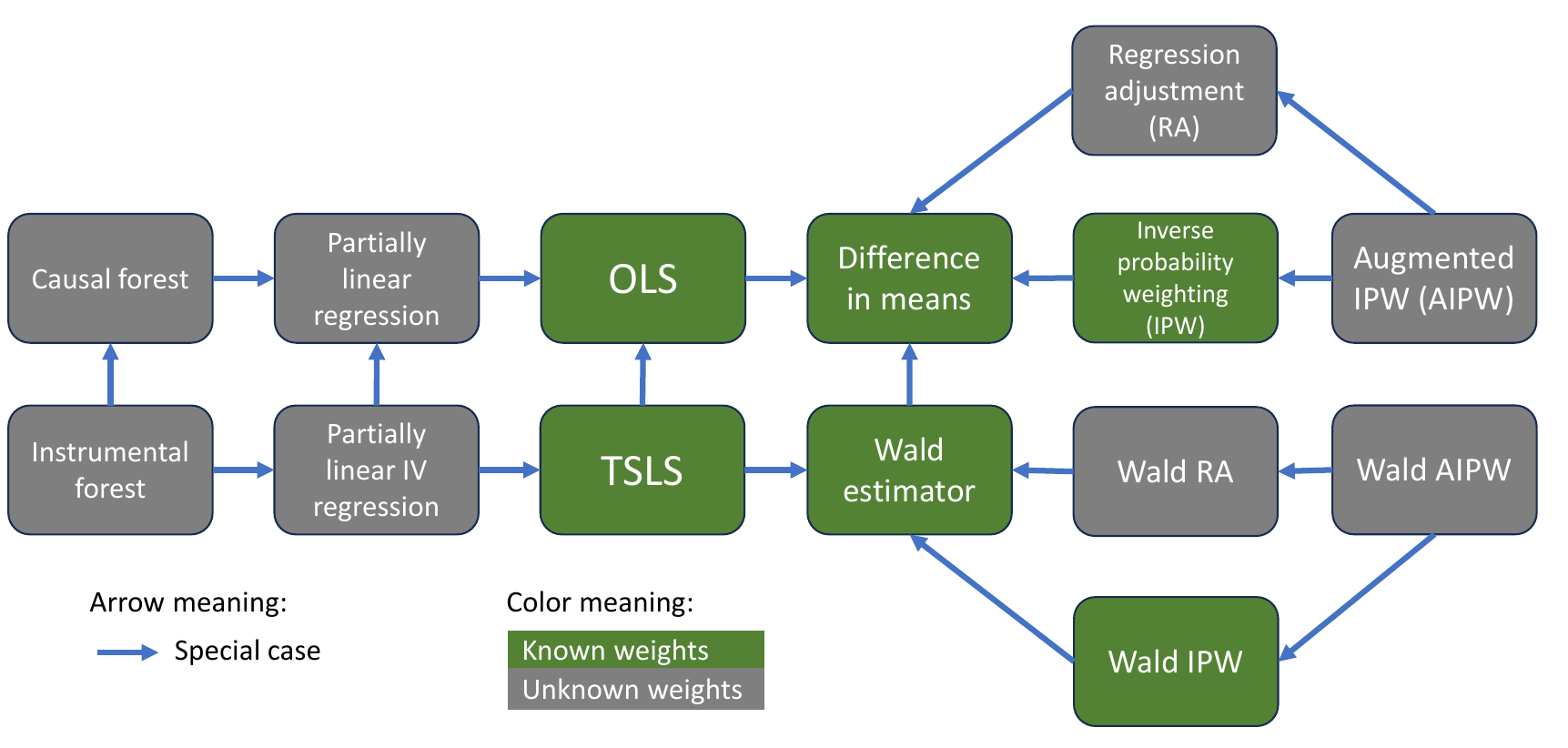}
\end{figure}

Figure \ref{fig:estimators} additionally illustrates how estimators are connected and for which outcome weights are already known in the literature.

\subsection{More on smoothers} \label{sec:app-smoothers}

Many common regression estimators admit a representation as smoother. We distinguish three classes of smoothers:\footnote{See for a similar discussion the recent literature using smoothers to explain properties of machine learning methods \cite{Curth2023ALearning,Curth2024WhySmoothers}.}
\begin{itemize}
\item $\bm{s}_i(X_i;\bm{X},\bm{Y},\epsilon_s)$ are \textit{smoothers} that may depend on the outcome vector and on any type of randomness in building the prediction model, e.g. by inducing randomness in a random forest and/or by cross-validating the hyperparameters.
\item $\bm{s}_i(X_i;\bm{X},\bm{Y},\epsilon_s) = \bm{s}_i(X_i;\bm{X},\bm{Y},\epsilon_s')~\forall~\epsilon_s \neq \epsilon_s'$ are \textit{deterministic smoothers} that do not depend on a random component, while still being outcome adaptive. One example would be (Post-)Lasso with data-driven penalty terms as implemented in the \texttt{hdm} R package of \citeA{Chernozhukov2016High-DimensionalR}.
\item $\bm{s}_i(X_i;\bm{X},\bm{Y},\epsilon_s) = \bm{s}_i(X_i;\bm{X},\bm{Y'},\epsilon_s')~\forall~\epsilon_s \neq \epsilon_s', \bm{Y} \neq \bm{Y'}$ are \textit{linear smoothers} and neither depend on the outcome vector nor on a random component. OLS specified without using the data is a canonical linear smoother but also kernel and series regressions with fixed tuning parameter are linear smoothers \cite<e.g.>{Stone1977ConsistentRegression,Buja1989LinearModels}.
\end{itemize}

The results in the main text merely require the existence of the smoother weights and do not depend on the smoother class. Therefore, we leave a more detailed discussion of the different classes for instances where the differences matter.

\subsection{\texttt{grf} package specific considerations}

The main text ignores some complications arising in the R package \texttt{grf} implementing causal/instrumental forests and AIPW. 

\subsubsection{Causal forest}\label{sec:app-grf-cf}

The first complication arises because the \texttt{grf} runs a weighted residual-on-residual regression \textit{with constant}. The coefficient of this constant is typically not exactly zero because the weighted residuals are not guaranteed to sum to zero. Therefore, implementing Equation \ref{eq:iw-if} will not exactly recover the package output. However, it can be achieved by defining the weighted least squares residual maker matrix $\bm{M^\alpha_{1_N}} := \bm{I_N} - \bm{1_N(1_N'diag(\bm{\alpha(\bm{x})})1_N)^{-1} 1_N'diag(\bm{\alpha(\bm{x})})}$ and using it in a modified version of \eqref{eq:iw-if} and for the causal forest as special case:
\begin{align} \label{eq:iw-if-aw}
\bm{\omega}^{if}(\bm{x})\bm{'} = (\bm{\hat{R}'}\bm{M^{\alpha^{if}}_{1_N}}diag(\bm{\alpha^{if}(\bm{x})})  \bm{\hat{V}})^{-1} \bm{\hat{R}'} \bm{M^{\alpha^{if}}_{1_N}} diag(\bm{\alpha^{if}(\bm{x})}) (\bm{I_N} - \bm{S}) \\
\bm{\omega}^{cf}(\bm{x})\bm{'} = (\bm{\hat{V}'}\bm{M^{\alpha^{cf}}_{1_N}}diag(\bm{\alpha^{cf}(\bm{x})})  \bm{\hat{V}})^{-1} \bm{\hat{V}'} \bm{M^{\alpha^{cf}}_{1_N}} diag(\bm{\alpha^{cf}(\bm{x})}) (\bm{I_N} - \bm{S})
\end{align}
This means we use a different pseudo-instrument compared to Equation \ref{eq:iw-if} but leave pseudo-outcome and -treatment unchanged. Consequently, the conclusion in Section \ref{sec:iw-probs-if} that causal/instrumental forests are scale-normalized unless C\ref{cond:osmatt}a is enforced remains valid because the pseudo-instrument does not affect the weights properties.

\subsubsection{AIPW}\label{sec:app-grf-aipw}

The second complication arises when estimating the average treatment effect using AIPW with the \texttt{average\_treatment\_effect()} function. As described in \citeA{Athey2019EstimatingApplication} Equation (8), the \texttt{grf} implementation applies an alternative representation of AIPW:
\begin{equation} \label{eq:mom-aipw-aw}
\mathbb{E}_N \Bigg[\underbrace{\hat{\tau}^{cf}(\bm{X_i}) + (\lambda_{1,i}^{ipw} -  \lambda_{0,i}^{ipw}) (Y_i - \hat{Y}_i - (D_i - \hat{D}_i)\hat{\tau}^{cf}(\bm{X_i}))}_{=:\tilde{Y}_i^{aipw-grf}} - \hat{\tau}^{aipw-grf} \Bigg] = 0
\end{equation}
It uses the CATE estimates obtained by the causal forest $\hat{\tau}^{cf}(\bm{X_i})$ as nuisance parameter and not the two separate outcome regressions. To derive the outcome weights we store the CATEs of every observation in $\bm{\hat{\tau}^{cf}} := (\hat{\tau}^{cf}(\bm{X_1}),...,\hat{\tau}^{cf}(\bm{X_N}))'$. The solution of \eqref{eq:mom-aipw-aw} in vector notation reads then
\begin{equation} \label{eq:aipw-4-aw}
\hat{\tau}^{aipw-grf} = N^{-1} \bm{1_N'}[ \bm{\hat{\tau}^{cf}} +  diag(\bm{\lambda_{1}^{ipw}} -  \bm{\lambda_{0}^{ipw}}) (\bm{Y} - \bm{\hat{Y}} - diag(\bm{D} - \bm{\hat{D}}) \bm{\hat{\tau}^{cf}}) ].
\end{equation}
We have established in Section \ref{sec:iw-if} how to get the $\bm{x}$-specific causal forest weights and store them in a CATE smoother matrix $\bm{S^\tau} := [\bm{\omega}^{cf}(\bm{X_1})~...~\bm{\omega}^{cf}(\bm{X_N})]'$ such that $\bm{\hat{\tau}^{cf}} = \bm{S^\tau Y}$. The AIPW weights of the \texttt{grf} implementation are then
\begin{align} \label{eq:iw-aipw-aw}
\bm{\omega^{aipw-grf'}} & = N^{-1} \bm{1_N'} [ \bm{S^\tau} + diag(\bm{\lambda_{1}^{ipw}} -  \bm{\lambda_{0}^{ipw}})  (\bm{I_N} - \bm{S} -  diag(\bm{D} - \bm{\hat{D}})\bm{S^\tau})]. 
\end{align}

The next step is to investigate under which conditions \texttt{grf}-AIPW is normalized where we use $\bm{\lambda^{ipw}} = \bm{\lambda_{1}^{ipw}} -  \bm{\lambda_{0}^{ipw}}$ for compactness:
\begin{align*}
    \bm{T^{aipw-grf} 1_N} 
    &= [\bm{S^\tau} + diag(\bm{\lambda^{ipw}})  (\bm{I_N} - \bm{S} -  diag(\bm{D} - \bm{\hat{D}})\bm{S^\tau})] \bm{1_N} \\
    &= \bm{S^\tau}  \bm{1_N} + \bm{\lambda^{ipw}} - diag(\bm{\lambda^{ipw}}) \bm{S}\bm{1_N} - diag(\bm{\lambda^{ipw}}) diag(\bm{D} - \bm{\hat{D}})\bm{S^\tau} \bm{1_N} \\
    \text{If C\ref{cond:affine-smooth}}  &= \bm{0_N} + \bm{\lambda^{ipw}} - \bm{\lambda^{ipw}} - diag(\bm{\lambda^{ipw}}) diag(\bm{D} - \bm{\hat{D}}) \bm{0_N}  = \bm{0_N} \Rightarrow \text{normalized}
\end{align*}
because we have shown in Section \ref{sec:iw-probs-if} that causal forest is normalized under C\ref{cond:affine-smooth} and therefore $\bm{S^\tau} \bm{1_N} = \bm{0_N}$. C\ref{cond:affine-smooth} holds in the implementation of \texttt{grf} by default because it applies a random forest to estimate the outcome prediction. However, this is not enough to guarantee fully-normalized weights:
\begin{align*}
    \bm{T^{aipw-grf} D} &= [\bm{S^\tau} + diag(\bm{\lambda^{ipw}})  (\bm{I_N} - \bm{S} -  diag(\bm{D} - \bm{\hat{D}})\bm{S^\tau})] \bm{D} \\
    &= \bm{S^\tau}  \bm{D} + \bm{\lambda_1^{ipw}} - diag(\bm{\lambda^{ipw}}) \bm{S}\bm{D} - diag(\bm{\lambda^{ipw}}) diag(\bm{D} - \bm{\hat{D}})\bm{S^\tau} \bm{D}  \\
    \text{If C\ref{cond:affine-smooth} and C\ref{cond:osmatt}a} & = \bm{1_N} + \bm{\lambda_1^{ipw}} - diag(\bm{\lambda^{ipw}}) \bm{\hat{D}} - diag(\bm{\lambda^{ipw}}) diag(\bm{D} - \bm{\hat{D}})\bm{1_N}  \\
    & = \bm{1_N} + \bm{\lambda_1^{ipw}} - diag(\bm{\lambda^{ipw}}) \bm{\hat{D}} - \bm{\lambda_1^{ipw}} + diag(\bm{\lambda^{ipw}}) \bm{\hat{D}} = \bm{1_N} = \bm{\tilde{D}^{aipw-grf}} \\
    &\Rightarrow \text{fully-normalized }
\end{align*}
This means that the AIPW estimator of \texttt{grf} is only scale-normalized because C\ref{cond:osmatt}a is not fulfilled by default. This is in contrast to other implementations that are self-fully-normalized as discussed in Section \ref{sec:aipw-prop}.

\subsection{Replicate Słoczyński et al. (2024) in the PIVE framework} \label{app:replicate-usw}

\citeA{Soczynski2024AbadiesEffect} consider in total five estimators. Their estimators $\hat{\tau}_t = \hat{\tau}_{a,1}$ correspond to the Wald-IPW without Condition \ref{cond:norm-ipw}b and $\hat{\tau}_{u}$ to the Wald-IPW with normalized weights and are already discussed in Section \ref{sec:wald-aipw-prop}. In addition the paper considers three estimators that require to define Abadie's \citeyear{Abadie2003SemiparametricModels} kappas:
\begin{itemize} \singlespacing
    \item $\kappa_0 := (1-D_i) \frac{(1-Z_i)-(1-\hat{D}_i)}{\hat{D}_i(1-\hat{D}_i)}$
    \item $\kappa_1 := D_i \frac{Z_i-\hat{D}_i}{\hat{D}_i(1-\hat{D}_i)}$
    \item $\kappa := 1 - D_i \frac{1 - Z_i}{1-\hat{D}_i} - (1-D_i) \frac{Z_i (1-\hat{D}_i)}{\hat{D}_i}$
\end{itemize}

The three estimators are now presented in their vector form:
\begin{itemize}
    \item $\hat{\tau}_{a} = (\bm{1_N'\kappa})^{-1}\bm{1_N'}diag(\bm{\kappa_1} - \bm{\kappa_0}) \bm{Y}$
    \item $\hat{\tau}_{a,0} = (\bm{1_N'\kappa_0})^{-1}\bm{1_N'}diag(\bm{\kappa_1} - \bm{\kappa_0}) \bm{Y}$
    \item $\hat{\tau}_{a,10} = (\bm{1_N'}\bm{1_N})^{-1}\bm{1_N'}diag\big(\bm{\kappa_1} (\bm{1_N'\kappa_1})^{-1} N  - \bm{\kappa_0} (\bm{1_N'\kappa_0})^{-1} N\big)  \bm{Y}$ 
\end{itemize}

Now we can apply the same strategies as in Section \ref{sec:ow-props-concrete} to replicate that $\hat{\tau}_{a}$ is fully-unnormalized, $\hat{\tau}_{a,0}$ is treated-unnormalized, and $\hat{\tau}_{a,10}$ is fully-normalized.

We note that $\hat{\tau}_{a}$ and $\hat{\tau}_{a,0}$ share the same transformation matrix and that the first two steps following shortcuts \ref{item:sc1} and \ref{item:sc2} are identical:
\begin{itemize}
    \item $diag(\bm{\kappa_1} - \bm{\kappa_0}) \bm{1_N} \neq \bm{0_N} \Rightarrow$ unnormalized
    \item $diag(\bm{\kappa_1} - \bm{\kappa_0}) \bm{D} = \bm{\kappa_1}$, which is not equal to the pseudo-treatments $\bm{\tilde{D}^{\hat{\tau}_{a}} } = \bm{\kappa}$ or $\bm{\tilde{D}^{\hat{\tau}_{a,0}} } = \bm{\kappa_0}$, respectively  $\Rightarrow$ not untreated-unnormalized
\end{itemize}
To finish the characterization, we check whether the transformation matrix applied to the untreated produces minus the pseudo-treatment (shortcut \ref{item:sc3}):
\begin{align*}
    diag(\bm{\kappa_1} - \bm{\kappa_0}) (\bm{1_N} - \bm{D}) & = -\bm{\kappa_0} \\
    \Rightarrow \bm{T^{\hat{\tau}_{a}}} (\bm{1_N} - \bm{D}) & \neq -\bm{\tilde{D}^{\hat{\tau}_{a}}}  \\
    \Rightarrow \bm{T^{\hat{\tau}_{a,0}}} (\bm{1_N} - \bm{D}) & = -\bm{\tilde{D}^{\hat{\tau}_{a,0}}} 
\end{align*}
We conclude that $\hat{\tau}_{a}$ is fully unnormalized, but that $\hat{\tau}_{a,0}$ is at least treated-unnormalized with untreated weights summing to minus one in line with \citeA{Soczynski2024AbadiesEffect}.

Finally, we investigate $\hat{\tau}_{a,10}$. Here, it does not suffice to focus on the transformation matrix but we have to consider the full numerator to check whether weights sum to zero
\begin{align*}
    \bm{1_N'} diag\big(\bm{\kappa_1} (\bm{1_N'\kappa_1})^{-1} N  - \bm{\kappa_0} (\bm{1_N'\kappa_0})^{-1} N\big)  \bm{1_N}
    & =     \bm{1_N'} \big(\bm{\kappa_1} (\bm{1_N'\kappa_1})^{-1} N  - \bm{\kappa_0} (\bm{1_N'\kappa_0})^{-1} N\big) \\
    & =     \bm{1_N'} \bm{\kappa_1} (\bm{1_N'\kappa_1})^{-1} N - \bm{1_N'}\bm{\kappa_0} (\bm{1_N'\kappa_0})^{-1} N \\
    &= N - N = 0 \Rightarrow \text{normalized}
\end{align*}
and even the full expression to see that they are in addition fully-normalized:
\begin{align*}
    (\bm{1_N'}\bm{1_N})^{-1}\bm{1_N'} diag\big(\bm{\kappa_1} (\bm{1_N'\kappa_1})^{-1} N  - \bm{\kappa_0} (\bm{1_N'\kappa_0})^{-1} N\big)  \bm{D}
    & = N^{-1} \bm{1_N'} \bm{\kappa_1} (\bm{1_N'\kappa_1})^{-1} N \\
    & = 1 \Rightarrow \text{fully-normalized}
\end{align*}

This final derivation highlights that the shortcuts \ref{item:sc1}-\ref{item:sc3} via the transformation matrix are only sufficient but not necessary to establish weights properties.

\subsection{More on outcome weights properties}

\subsubsection{Graphical illustration of the classes}

Figure \ref{fig:weights-classes} illustrates the different classes formally defined in Table \ref{tab:weights-class}:

\begin{figure}[ht]
    \centering
    \caption{Outcome weights classes}
    \includegraphics[width=0.5\linewidth]{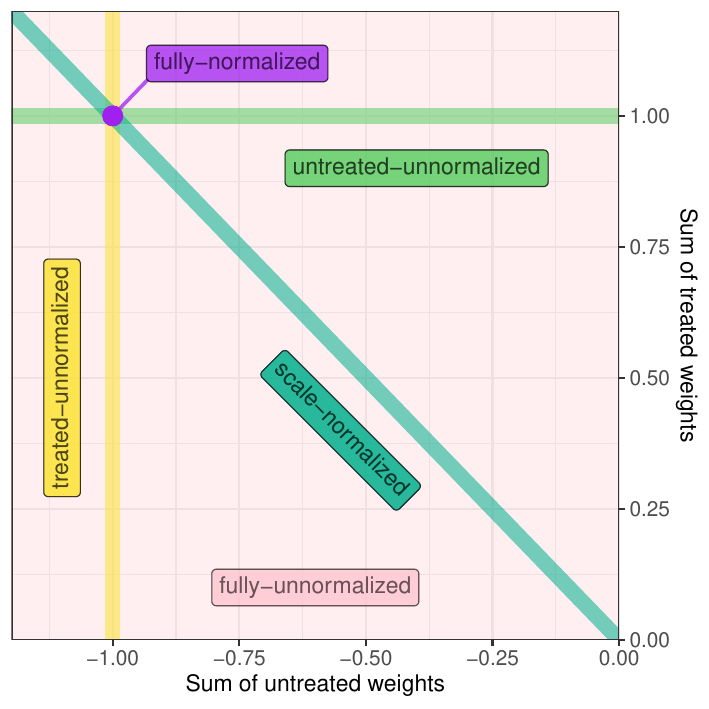}    
    \label{fig:weights-classes}
\end{figure}

\subsubsection{Full results for weights properties}

Table \ref{tab:app-iw-props} is an expanded version of Table \ref{tab:iw-props} adding columns two, four and five. The second column stores which conditions are (not) fulfilled by construction for particular estimators, which influences the properties. The fourth and fifth column collect when estimators and (un)treated-normalized without being also normalized at the same time. These require rather artificial implementation decisions. For example in column four using affine smoothers only for the untreated but not for the treated outcome prediction ensures treated-unnormalized weights.

\begin{center}
\begin{landscape}
\begin{threeparttable}[ht] 
\onehalfspacing
\caption{Conditions for closed-form and properties of outcome weights} \label{tab:app-iw-props}
\begin{tabular}{lccccccc}   
\toprule
\textbf{Estimator} & \textbf{By construction} & \textbf{Closed-form} & \textbf{Treated-unnorm.} &  \textbf{Untreated-unnorm.} & \textbf{Normalized} & \textbf{Fully-normalized}  \\ 
\midrule
IF  & - & C\ref{cond:smoothers}a & - & C\ref{cond:osmatt}a & C\ref{cond:affine-smooth} & C\ref{cond:affine-smooth} \& C\ref{cond:osmatt}a  \\ 
PLR-IV  & - & C\ref{cond:smoothers}a & - & C\ref{cond:osmatt}a & C\ref{cond:affine-smooth} & C\ref{cond:affine-smooth} \& C\ref{cond:osmatt}a  \\ 
TSLS & C\ref{cond:smoothers}a \& C\ref{cond:osmatt}a & \cmark & - & \cmark & C\ref{cond:ols-constant} & C\ref{cond:ols-constant} \\ 
Wald & C\ref{cond:smoothers}a \& C\ref{cond:affine-smooth} \& C\ref{cond:osmatt}b & \cmark & - & \cmark & \cmark & \cmark \\ 
CF & - & C\ref{cond:smoothers}a & - & C\ref{cond:osmatt}a & C\ref{cond:affine-smooth} & C\ref{cond:affine-smooth} \& C\ref{cond:osmatt}a  \\
PLR & - & C\ref{cond:smoothers}a & - & C\ref{cond:osmatt}a & C\ref{cond:affine-smooth} & C\ref{cond:affine-smooth} \& C\ref{cond:osmatt}a  \\
OLS & C\ref{cond:smoothers}a \& C\ref{cond:osmatt}a & \cmark & - & \cmark & C\ref{cond:ols-constant} & C\ref{cond:ols-constant} \\ 
DiM & C\ref{cond:smoothers}a \& C\ref{cond:affine-smooth} \& C\ref{cond:osmatt}a & \cmark & - & - & \cmark & \cmark \\ 
AIPW & - & C\ref{cond:smoothers}b & $\bm{S_0^d 1_N} = \bm{1_N}$ & $\bm{S_1^d 1_N} = \bm{1_N}$ & C\ref{cond:affine-smooth} & C\ref{cond:affine-smooth} \& C\ref{cond:nsbg}  \\
RA & - & C\ref{cond:smoothers}b & $\bm{S_0^d 1_N} = \bm{1_N}$ & $\bm{S_1^d 1_N} = \bm{1_N}$ & C\ref{cond:affine-smooth} & C\ref{cond:affine-smooth} \& C\ref{cond:nsbg}  \\ 
IPW & C\ref{cond:smoothers}b \& C\ref{cond:nsbg} \& \cancel{C\ref{cond:affine-smooth}} & \cmark & with $\bm{\lambda_{0}^{norm}}$ & with $\bm{\lambda_{1}^{norm}}$ & C\ref{cond:norm-ipw}a & C\ref{cond:norm-ipw}a \\ 
Wald-AIPW & - & C\ref{cond:smoothers}c & -  & C\ref{cond:osmatt}b & C\ref{cond:affine-smooth} & C\ref{cond:affine-smooth} \& C\ref{cond:osmatt}b  \\
Wald-RA & - & C\ref{cond:smoothers}c & - & C\ref{cond:osmatt}b & C\ref{cond:affine-smooth} & C\ref{cond:affine-smooth} \& C\ref{cond:osmatt}b  \\
Wald-IPW & C\ref{cond:smoothers}c \& C\ref{cond:osmatt}b \& \cancel{C\ref{cond:affine-smooth}} & \cmark & - & \cmark & C\ref{cond:norm-ipw}b & C\ref{cond:norm-ipw}b \\ 
\bottomrule
\end{tabular}
\begin{tablenotes} \item \textit{Notes:} A ``-'' in columns 4 and 5 indicates that no condition of Section \ref{sec:imp-detail} leads  to (un)treated-unnormalized weights without making them also normalized. \end{tablenotes}  
\end{threeparttable}   
\end{landscape}    
\end{center}

\end{appendices}

\end{document}